\documentclass{new_tlp}

\usepackage[utf8]{inputenc}
\usepackage{mathptmx}
\usepackage[lighttt]{lmodern} % required to get (visibly) bold typewriter fonts
\usepackage{microtype}
\usepackage{amsmath,amssymb}
\usepackage{url}
\usepackage{graphicx}
\usepackage{listings}
\usepackage{caption}
\usepackage{footmisc}
\newtheorem{example}{Example}
\newtheorem{theorem}{Theorem}

\newcommand{\eofex}{\hfill$\blacksquare$}

\newcommand{\astep}{$\forall$-step}
\newcommand{\estep}{$\exists$-step}
\newcommand{\vars}{\ensuremath{\mathcal{F}}}
\newcommand{\stat}{\ensuremath{s}}
\newcommand{\init}{\ensuremath{\stat_0}}
\newcommand{\goal}{\ensuremath{\stat_\star}}
\newcommand{\acts}{\ensuremath{\mathcal{O}}}
\newcommand{\var}{\ensuremath{x}}
\newcommand{\dom}[1]{\ensuremath{#1^d}}
\newcommand{\act}{\ensuremath{a}}
\newcommand{\val}{\ensuremath{v}}
\newcommand{\pred}[1]{\ensuremath{#1^c}}
\newcommand{\post}[1]{\ensuremath{#1^e}}
\newcommand{\occur}[2]{\ensuremath{o(#1,#2)}}
\newcommand{\actp}{\ensuremath{A}}
\newcommand{\defi}[1]{\tilde{#1}}
\newcommand{\defg}{\defi{\stat}_\star}

\newcommand{\fact}{\ensuremath{I}}
\newcommand{\base}{\ensuremath{B}}
\newcommand{\query}[1]{\ensuremath{Q(#1)}}
\newcommand{\step}[1]{\ensuremath{S(#1)}}
\newcommand{\steps}[1]{\ensuremath{S^s(#1)}}

\newcommand{\stepa}[1]{\ensuremath{S^\forall(#1)}}
\newcommand{\stepe}[1]{\ensuremath{S^\exists(#1)}}
\newcommand{\stepc}[1]{\ensuremath{S^E(#1)}}
\newcommand{\stepr}[1]{\ensuremath{S^R(#1)}}
\newcommand{\stepp}[1]{\ensuremath{S^p(#1)}}
\newcommand{\smod}{\ensuremath{M}}

\newcommand{\sysfont}{\textit}
\newcommand{\plasp}{\sysfont{plasp}}
\newcommand{\iclingo}{\sysfont{iclingo}}
\newcommand{\clingo}{\sysfont{clingo}}
\newcommand{\fastdownward}{\sysfont{Fast Downward}}
\newcommand{\madagascar}{\sysfont{Madagascar}}

\newcommand{\queryatom}[1]{\ensuremath{\{\text{\textup{\lstinline{query(}}}#1\text{\textup{\lstinline{).}}}\}}} % {\ensuremath{\{\texttt{query(}#1\texttt{)}\}}}

\newcommand{\algfont}{\textsf}
\newcommand{\alga}[2]{\algfont{A}\ensuremath{#1#2}}
\newcommand{\algb}[2]{\algfont{B}\ensuremath{#1#2}}

\newcommand{\algs}[2]{\algfont{S}\ensuremath{#1#2}}
\newcommand{\encs}{\ensuremath{^s}}
\newcommand{\enca}{\ensuremath{^\forall}}
\newcommand{\ence}{\ensuremath{^\exists}}
\newcommand{\encc}{\ensuremath{^E}}
\newcommand{\encr}{\ensuremath{^R}}
\newcommand{\encg}{\ensuremath{^G}}
\newcommand{\heu}{\ensuremath{_p}}

\newcommand{\mad}{\algfont{M}}
\newcommand{\madp}{\mad\heu}

\begin{document}

\title{\plasp~3: Towards Effective ASP Planning}

\author[Y. Dimopoulos, M. Gebser, P. Lühne, J. Romero, and T. Schaub]{
  Yannis Dimopoulos
  \\
  University of Cyprus
  \and
  Martin Gebser
  \\
  University of Klagenfurt, % Austria, 
  Graz University of Technology, % Austria, 
  and 
  University of Potsdam % , Germany
  \and
  Patrick Lühne
  \
  Javier Romero
  \\
  University of Potsdam % , Germany
  \and
  Torsten Schaub
  \\
  INRIA Rennes % , France, 
  and
  University of Potsdam % , Germany
}

\submitted{[n/a]}
\revised{[n/a]}
\accepted{[n/a]}

\maketitle

\begin{abstract}
% Automated planning is one of the earliest application areas of ASP
% and forms the basis of reasoning about actions and change.
% Since then, however, the major emphasis in ASP laid on capturing various aspects of action-driven reasoning
% rather than its effective implementation.
% We address this shortcoming with the new \plasp{} system,
% originally designed as a PDDL-to-ASP translator.
% The new version extends its predecessor in several ways.
We describe the new version of the PDDL-to-ASP translator \plasp{}.
First, it  widens the range of accepted PDDL features.
Second, it contains novel planning encodings,
some inspired by SAT planning and others exploiting ASP features such as well-foundedness. % reachability.
All of them are designed for handling multivalued fluents in order to capture both PDDL as well as SAS planning formats.
Third,
enabled by multishot ASP solving,
it offers advanced planning algorithms also borrowed from SAT planning.
As a result,
\plasp{} provides us with an ASP-based framework for studying a variety of planning techniques in a uniform setting.
Finally,
we demonstrate in an empirical analysis that these techniques have a significant impact on the performance of ASP planning.

\medskip\noindent
{\em Under consideration for publication in Theory and Practice of Logic Programming (TPLP)}
\end{abstract}

%%% Local Variables: 
%%% mode: latex
%%% TeX-master: "paper"
%%% End: 

\section{Introduction}\label{sec:introduction}

Reasoning about actions and change constitutes a major challenge to any formalism for knowledge representation and reasoning.
It therefore comes as no surprise that Automated Planning \cite{dineko97a} % ,lifschitz02a
was among the first substantial applications of Answer Set Programming (ASP; \cite{lifschitz02a}). % baral02a
Meanwhile this has led to manifold action languages \cite{gellif98a}, % ,eifalepfpo03b,leliya13a
various applications in dynamic domains \cite{bargel00a}, % ,nobagewaba01a,bachtrtrjobe04a,erakpa12a,khyalelist14a
but only few adaptions of Automated Planning techniques \cite{sobanamc03a}.
Although such approaches have provided us with diverse insights into how relevant concepts are expressed in ASP,
almost no attention has been paid to making reasoning about actions and change effective.
This is insofar surprising as a lot of work has been dedicated to planning with techniques from the area of Satisfiability Testing (SAT; \cite{SATHandbook}),
a field often serving as a role model for ASP.

\begin{figure}[t]
\centering
\includegraphics{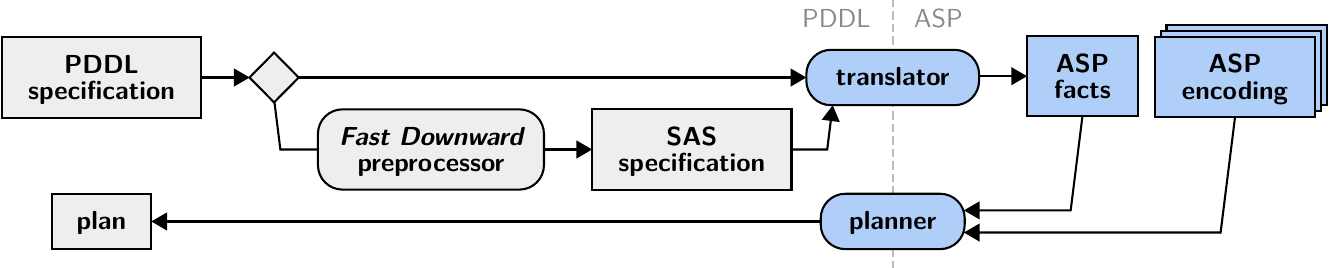}
\caption{Solving PDDL inputs with \plasp's workflow (highlighted in blue)\label{fig:plasp}}
\end{figure}

We address this shortcoming with the third series of the \plasp{} system.
From its inception,
the purpose of \plasp{} was to provide an elaboration-tolerant platform to planning by using ASP.
Already its original design \cite{gekaknsc11a} foresaw 
to compile planning problems formulated in the Planning Domain Definition Language (PDDL; \cite{mcdermott98a}) into ASP facts and 
to use ASP metaencodings for modeling alternative planning techniques.
These could then be solved with fixed horizons (and optimization) or in an incremental fashion. % (with \iclingo\ \cite{gekakaosscth08a}).
The redesigned \plasp~3 system processes planning problems specified in PDDL
according to the workflow visualized in Figure~\ref{fig:plasp}.
% The redesigned \plasp~3 system features
At the beginning, a PDDL input may be subject to
optional preprocessing by the state-of-the-art planning system \fastdownward\ \cite{helmert06a} via the intermediate SAS format.
The translator component of \plasp~3 otherwise performs
a normalization step to transform complex PDDL expressions into a simplified core format,
which results in 
a homogeneous factual representation capturing both PDDL and SAS inputs (with multivalued fluents).
% and a normalization step to support advanced PDDL features. % such as nested expressions in preconditions as well as existential and universal quantifiers.
%
Moreover,
\plasp~3 provides a spectrum of ASP encodings ranging 
from adaptions of known SAT encodings \cite{riheni06a}
to novel encodings taking advantage of ASP-specific concepts.
Finally, the planner component of
\plasp~3 offers sophisticated planning algorithms, also stemming from SAT planning \cite{riheni06a},
by taking advantage of multishot ASP solving.
Given the common structure of various incremental ASP encodings, % makes
\plasp's planning framework is also applicable to dynamic domains beyond PDDL, e.g.,
the planner could be run on ASP encodings of finite model finding
\cite{gesasc11a}
instead of planning.%
%
% The usual workflow of \plasp~3, though, is summarized in Figure~\ref{fig:plasp}.
%
%
% In what follows,
% we presuppose some basic familiarity with ASP and refer the reader for details to the literature.

The outline of this paper is as follows.
In Section~\ref{sec:encodings}, we introduce STRIPS-like planning tasks
and devise ASP encodings for sequential as well as a range of parallel representations of plans.
Section~\ref{sec:planner} is dedicated to the planner component of \plasp~3,
presenting the planning algorithms, guess-and-check strategies, and the planning heuristic it supports.
In Section~\ref{sec:pddl:asp},
we turn to the functionalities provided by \plasp's translator,
including normalization for dealing with advanced PDDL features
and the handling of constructs comprised in the intermediate SAS format.
Section~\ref{sec:experiments} reports about our experiments, empirically
evaluating the devised encodings and planning algorithms on PDDL inputs
as well as ASP planning benchmarks.
Finally, Section~\ref{sec:discussion} concludes the paper with a summary
of the achieved results and future work.
This paper extends a previous conference version \cite{digelurosc17a},
which did not include the guess-and-check strategies presented in Section~\ref{sec:planner},
the description of the translator component of \plasp~3 given in Section~\ref{sec:pddl:asp}, and
experimental results on the benchmark set by \cite{rintanen12a} as well as ASP planning benchmarks.

%%% Local Variables: 
%%% mode: latex
%%% TeX-master: "paper"
%%% End: 

\section{ASP Encodings for Planning}
\label{sec:encodings}

We consider STRIPS-like (multivalued) \emph{planning tasks} according to \cite{helmert06a},
given by a 4-tuple % $\ptask = 
$\langle \vars,\init,\goal,\acts \rangle$, in which
\begin{itemize}
\item $\vars$ is a finite set of state variables, also called \emph{fluents},
      where each $\var\in\vars$ has an associated finite domain $\dom{\var}$ of possible values for~$\var$,
\item $\init$ is a \emph{state}, i.e., a (total) function such that $\init(\var)\in\dom{\var}$
      for each $\var\in\vars$,
\item $\goal$ is a \emph{partial state} (listing goal conditions), i.e., a function such that $\goal(\var)\in\dom{\var}$      
      % whenever $\goal(\var)$ is defined for some $\var\in\vars$,
      for each $\var\in\defg$, where $\defg$ denotes the set of all $\var\in\vars$
      such that $\goal(\var)$ is defined,
      and
\item $\acts$ is a finite set of operators, also called \emph{actions},
      where $\pred{\act}$ and $\post{\act}$ in $\act=\langle \pred{\act},\post{\act} \rangle$ are
      partial states
      denoting the \emph{precondition} and \emph{postcondition} of~$\act$ for each $\act\in\acts$.
\end{itemize}
Given a state $\stat$ and an action $\act\in\acts$,
the \emph{successor state} $\occur{\act}{\stat}$ obtained by applying $\act=\langle \pred{\act},\post{\act} \rangle$ in~$\stat$
is defined if $\pred{\act}(\var)=\stat(\var)$ for each $\var\in\pred{\defi{\act}}$,
% $\var\in\vars$ such that $\pred{\act}(\var)$ is defined,
and undefined otherwise.
Provided that $\stat'=\occur{\act}{\stat}$ is defined,
$\stat'(\var)=\post{\act}(\var)$ for each $\var\in\post{\defi{\act}}$,
% $\var\in\vars$ such that $\post{\act}(\var)$ is defined,
and $\stat'(\var)=\stat(\var)$ for each $\var\in\vars\setminus\post{\defi{\act}}$. % otherwise.
That is, if the successor state $\occur{\act}{\stat}$ is defined,
it includes the postcondition of~$\act$ and keeps any other fluents unchanged from~$\stat$.
We extend the notion of a successor state to sequences $\langle \act_1,\dots,\act_n \rangle$
of actions by letting
$\occur{\langle \act_1,\dots,\act_n \rangle}{\stat}=\occur{\act_n}{\occur{\dots}{\occur{\act_1}{\stat}\dots}}$,
provided that
$\occur{\act_i}{\occur{\dots}{\occur{\act_1}{\stat}\dots}}$ is defined
for all $1\leq i\leq n$.
Given this, a \emph{sequential plan} is a sequence $\langle \act_1,\dots,\act_n \rangle$
of actions such that $\stat'=\occur{\langle \act_1,\dots,\act_n \rangle}{\init}$ is defined
and $\stat'(\var)=\goal(\var)$ for each $\var\in\defg$,
i.e., the goal conditions specified by $\goal$ have to hold in $\stat'$.
% whenever $\goal(\var)$ is defined for some $\var\in\vars$.

Several \emph{parallel} representations of sequential plans have been investigated
in the literature \cite{dineko97a,riheni06a,wehrin07a}.
We call a set $\{ \act_1,\dots,\act_k \}\subseteq\acts$ of actions \emph{confluent} if
$\post{\act_i}(\var)=\post{\act_j}(\var)$
% whenever $\post{\act_i}(\var)$ and $\post{\act_j}(\var)$ are both defined for some $\var\in\vars$
for all $1\leq i<j\leq k$ and each $\var\in\post{\defi{\act}_i}\cap\post{\defi{\act}_j}$.
Given a state $\stat$ and a confluent set $\actp=\{ \act_1,\dots,\act_k \}$ of actions, $\actp$ is 
\begin{itemize}
\item \emph{\astep\ serializable} in~$\stat$ if
      $\occur{\langle \act_1',\dots,\act_k' \rangle}{\stat}$
      is defined for any sequence $\langle \act_1',\linebreak[1]\dots,\linebreak[1]\act_k' \rangle$
      such that $\{\act_1',\dots,\act_k'\}=\actp$;
\item \emph{\estep\ serializable} in~$\stat$ if
      $\pred{\act}(\var)=\stat(\var)$,
      for each $\act\in\actp$ and $\var\in\pred{\defi{\act}}$, % $\var\in\vars$ such that $\pred{\act}(\var)$ is defined,
      and $\occur{\langle \act_1',\linebreak[1]\dots,\linebreak[1]\act_k' \rangle}{\stat}$ is defined
      for some sequence $\langle \act_1',\linebreak[1]\dots,\linebreak[1]\act_k' \rangle$
      such that $\{\act_1',\dots,\act_k'\}=\actp$;
\item \emph{relaxed \estep\ serializable} in~$\stat$ if
      $\occur{\langle \act_1',\dots,\act_k' \rangle}{\stat}$ is defined
      for some sequence $\langle \act_1',\linebreak[1]\dots,\linebreak[1]\act_k' \rangle$
      such that $\{\act_1',\dots,\act_k'\}=\actp$.
\end{itemize}
Note that any \astep\ serializable set~$\actp$ of actions is likewise \estep\ serializable,
and similarly any \estep\ serializable~$\actp$ is relaxed \estep\ serializable.
In particular, the condition that any sequence built from a \astep\ serializable~$\actp$
leads to a (defined) successor state implies that the precondition of each action in~$\actp$
must already be established in~$\stat$, which is also required for \estep\ serializable sets,
but not for relaxed \estep\ serializable sets.
We extend the three serialization concepts to plans by calling a sequence
$\langle \actp_1,\dots,\actp_m \rangle$ of confluent sets of actions
a \emph{\astep}, \emph{\estep}, or \emph{relaxed \estep\ plan} if
$\stat_m(\var)=\goal(\var)$, for each $\var\in\defg$, % , whenever $\goal(\var)$ is defined for some $\var\in\vars$,
and each % set~$\actp_i$ of actions 
$\actp_i$ is \astep, \estep, or relaxed \estep\ serializable, respectively, in $\stat_{i-1}$
for $1\leq i\leq m$,
where $\stat_i(\var)=\post{\act}(\var)$ for each $\act\in\actp_i$ and $\var\in\post{\defi{\act}}$,
% $\var\in\vars$ such that there is some $\act\in\actp_i$ for which $\var\in\post{\defi{\act}}$,
% $\post{\act}(\var)$ is defined,
and $\stat_i(\var)=\stat_{i-1}(\var)$ for each $\var\in\vars\setminus\bigcup_{\act\in\actp_i}\post{\defi{\act}}$. % otherwise.
That is, parallel representations partition some sequential plan
such that each part~$\actp_i$ is \astep, \estep, or relaxed \estep\ serializable in the state
obtained by applying the actions preceding~$\actp_i$.
Also note that, in case of (relaxed) \estep\ plans, the confluence requirement % of sets of actions
achieves tractability of deciding whether a set of actions is (relaxed) \estep\ serializable,
which becomes NP-hard otherwise \cite{riheni06a}.

\begin{example}\label{ex:parallel}
Consider a planning task $\langle \vars,\init,\goal,\acts \rangle$ with
$\vars = \{\var_1,\var_2,\var_3,\var_4,\var_5\}$
such that $\dom{\var_1}=\dom{\var_2}=\dom{\var_3}=\dom{\var_4}=\dom{\var_5}=\{0,1\}$,
$\init = \{\var_1=0,\linebreak[1]\var_2=0,\linebreak[1]\var_3=0,\linebreak[1]\var_4=0,\linebreak[1]\var_5=0\}$,
$\goal = \{\var_4=1,\var_5=1\}$, and
$\acts = \{\act_1,\linebreak[1]\act_2,\linebreak[1]\act_3,\linebreak[1]\act_4\}$,
where
$\act_1=\langle \{\var_1=\nolinebreak 0\},\linebreak[1]\{\var_1=1,\linebreak[1]\var_2=1\} \rangle$,
$\act_2=\langle \{\var_3=0\},\linebreak[1]\{\var_1=1,\linebreak[1]\var_3=1\} \rangle$,
$\act_3=\langle \{\var_2=1,\linebreak[1]\var_3=1\},\linebreak[1]\{\var_4=1\} \rangle$, and
$\act_4=\langle \{\var_2=1,\linebreak[1]\var_3=1\},\linebreak[1]\{\var_5=1\} \rangle$.
One can check that
$\langle \act_1,\act_2,\act_3,\act_4\rangle$ and
$\langle \act_1,\act_2,\act_4,\act_3\rangle$ are the two sequential plans
consisting of four actions.
The \astep\ plan with fewest sets of actions is given by
$\langle \{\act_1\},\linebreak[1]\{\act_2\},\linebreak[1]\{\act_3,\act_4\} \rangle$.
Similarly,
$\langle \{\act_1,\act_2\},\linebreak[1]\{\act_3,\act_4\} \rangle$ is the \estep\ plan
with fewest sets of actions.
Finally, the relaxed \estep\ plan 
$\langle \{\act_1,\act_2,\act_3,\act_4\} \rangle$ 
consists of one set of actions only.
\eofex
\end{example}

\begin{lstlisting}[numbers=none,float,label=lst:facts,caption={ASP fact representation of the planning task from Example~\ref{ex:parallel}}]
fluent(x1).  fluent(x2).  fluent(x3).  fluent(x4).  fluent(x5).
value(x1,0). value(x2,0). value(x3,0). value(x4,0). value(x5,0).
value(x1,1). value(x2,1). value(x3,1). value(x4,1). value(x5,1).

init(x1,0).  init(x2,0).  init(x3,0).  init(x4,0).  init(x5,0).
                                       goal(x4,1).  goal(x5,1).

action(a1).     action(a2).     action(a3).     action(a4).
prec(a1,x1,0).  prec(a2,x3,0).  prec(a3,x2,1).  prec(a4,x2,1).
post(a1,x1,1).  post(a2,x1,1).  prec(a3,x3,1).  prec(a4,x3,1).
post(a1,x2,1).  post(a2,x3,1).  post(a3,x4,1).  post(a4,x5,1).
\end{lstlisting}
\begin{lstlisting}[label=lst:baseline,caption={Common part of sequential and parallel % ASP 
encodings for STRIPS-like planning},float,lastline=15]
holds(X,V,0) :- init(X,V).

#program check(t).

:- query(t), goal(X,V), not holds(X,V,t).

#program step(t).

{holds(X,V,t) : value(X,V)} = 1 :- fluent(X).

{occurs(A,t)} :- action(A).

:- occurs(A,t), post(A,X,V), not holds(X,V,t).

:- holds(X,V,t), not holds(X,V,t-1), not occurs(A,t) : post(A,X,V).
\end{lstlisting}
In ASP, we represent a planning task like the one from Example~\ref{ex:parallel}
by facts as given in Listing~\ref{lst:facts}.
The facts can then be combined with encodings such that stable models
correspond to sequential, \astep, \estep, or relaxed \estep\ plans.
The rules as well as integrity constraints in Listing~\ref{lst:baseline} form the common core of
respective incremental encodings \cite{gekakasc14b} and are grouped
into three parts:
a subprogram \lstinline{base}, including the rule in Line~1, which is not preceded by
any \lstinline{#program} directive;
a parameterized subprogram \lstinline{check(t)},
containing the integrity constraint in Line~5,
in which the parameter~\lstinline{t} serves as placeholder for
successive integers starting from~\lstinline{0}; and
a parameterized subprogram \lstinline{step(t)},
comprising the rules and integrity constraints below the \lstinline{#program} directive in Line~7,
whose parameter~\lstinline{t} stands for successive integers starting from~\lstinline{1}.
By first instantiating the \lstinline{base} subprogram along with \lstinline{check(t)},
where~\lstinline{t} is replaced by~\lstinline{0},      and then proceeding with
% \comment{JR: This reads a bit complicated, maybe move ``once'' to ``By instantiating once''?}
integers from~\lstinline{1} for~\lstinline{t} in \lstinline{check(t)} and \lstinline{step(t)},
an incremental encoding can be gradually unrolled.
We take advantage of this to capture plans of increasing length,
expressed by the latest integer used to replace~\lstinline{t} with.

In more detail, the rule in Line~1 of Listing~\ref{lst:baseline}
maps facts specifying $\init$ to atoms over the predicate \lstinline{holds}/3,
in which the third argument \lstinline{0} refers to the initial state.
Starting from~\lstinline{0} for the parameter~\lstinline{t},
the integrity constraint in Line~5 then tests whether the
conditions of~$\goal$ are established,
where the dedicated atom \lstinline{query(t)} is set to true only for
the latest integer taken for~\lstinline{t}.
%
% \comment{JR: Maybe say ``externally set to true'', or some hint to better understand what is this setting to true.}
%
This allows for increasing the plan length by successively instantiating
the subprograms \lstinline{check(t)} and \lstinline{step(t)} with further integers.
The latter subprogram includes the choice rule in Line~9 to generate a successor state
such that each fluent $\var\in\vars$ is mapped to some value in its domain~$\dom{\var}$.
The other choice rule in Line~11 permits to unconditionally pick actions to apply,
expressed by atoms over \lstinline{occurs}/2,
in order to obtain a corresponding successor state.
Given that both sequential and parallel plans are such that the postcondition of an applied action
holds in the successor state, 
the integrity constraint in Line~13 asserts the respective postcondition(s),
which guarantees the confluence of any
set of actions to be applied in parallel.
On the other hand, fluents unaffected by applied actions must remain unchanged,
% which is reflected by the rules in Lines~15 and~16 along with the integrity constraint in Line~17,
and the integrity constraint in Line~15
% restricting changed fluents to postconditions of applied actions. 
thus requires changed fluents to be established by means of applied actions.%
\footnote{%
The variables \lstinline[basicstyle=\ttfamily\footnotesize]{X} and \lstinline[basicstyle=\ttfamily\footnotesize]{V}
occur outside the scope of \emph{conditional literals},
composed by the `\lstinline[basicstyle=\ttfamily\footnotesize]{:}' connective,
and are thus \emph{global} in Line~15,
while \lstinline[basicstyle=\ttfamily\footnotesize]{A} is \emph{local} to % the conditional literal
`\lstinline[basicstyle=\ttfamily\footnotesize]{not occurs(A,t) : post(A,X,V)}'.
An according instantiation of the integrity constraint in Line~15,
taking the values \lstinline[basicstyle=\ttfamily\footnotesize]{x1} and \lstinline[basicstyle=\ttfamily\footnotesize]{1}
for \lstinline[basicstyle=\ttfamily\footnotesize]{X} and \lstinline[basicstyle=\ttfamily\footnotesize]{V}
relative to the facts in Listing~\ref{lst:facts}, is
`\lstinline[basicstyle=\ttfamily\footnotesize]{:- holds(x1,1,t),}
 \lstinline[basicstyle=\ttfamily\footnotesize]{not holds(x1,1,t-1),}
 \lstinline[basicstyle=\ttfamily\footnotesize]{not occurs(a1,t),}
 \lstinline[basicstyle=\ttfamily\footnotesize]{not occurs(a2,t).}'
Note that the actions \lstinline[basicstyle=\ttfamily\footnotesize]{a1} and
\lstinline[basicstyle=\ttfamily\footnotesize]{a2},
which have $\var_1=1$ in their postconditions, are taken as values for the local variable
\lstinline[basicstyle=\ttfamily\footnotesize]{A}, where the resulting instances of the literal
`\lstinline[basicstyle=\ttfamily\footnotesize]{not occurs(A,t)}'
on the left-hand side of `\lstinline[basicstyle=\ttfamily\footnotesize]{:}'
are connected conjunctively in a propositional rule with the
constant~\lstinline[basicstyle=\ttfamily\footnotesize]{t} as placeholder for integers.
For a detailed account of the language of the ASP system \clingo,
we refer the reader to \cite{PotasscoUserGuide}.}

\begin{lstlisting}[label=lst:sequential,caption={Extension of Listing~\ref{lst:baseline} for encoding sequential plans},float,firstnumber=17,firstline=3]
:- occurs(A,t), prec(A,X,V), not holds(X,V,t-1).

:- #count{A : occurs(A,t)} > 1.
\end{lstlisting}
The common encoding part described so far takes care of matching
successor states to postconditions of applied actions,
while requirements regarding preconditions are subject to the kind of plan
under consideration and expressed by dedicated additions to the 
\lstinline{step(t)} subprogram.
To begin with, the two integrity constraints added in Listing~\ref{lst:sequential}
address \emph{sequential plans} by, in Line~17, asserting the precondition of an applied action
to hold in the state referred to by \lstinline{t-1}
and, in Line~19,
denying multiple actions to be applied in parallel.
Note that, if the plan length or the latest integer taken for~\lstinline{t}, respectively,
exceeds the minimum number of actions required to establish the conditions of~$\goal$,
the encoding of sequential plans given by Listings~\ref{lst:baseline} and~\ref{lst:sequential}
permits idle states in which no action is applied.
While      idle states cannot emerge when using the basic \iclingo\ control loop \cite{gekakasc14b}
% of \clingo\
to compute shortest plans,
they are essential for the planner presented in Section~\ref{sec:planner}
in order to increase the plan length in more flexible ways.

Turning to parallel representations, Listing~\ref{lst:forall} shows additions
dedicated to \emph{\astep\ plans},
where the integrity constraint in Line~17 is the same as in Listing~\ref{lst:sequential} before.
This guarantees the preconditions of applied actions to hold,
while their confluence is already taken care of by means of the
integrity constraint in Line~13 of Listing~\ref{lst:baseline}.
It thus remains to make sure that applied actions do not interfere
in any way that would disable a serialization,
which essentially means that the precondition of an applied action~$\act$
must not be invalidated by another action applied in parallel.
For a fluent $\var\in\pred{\defi{\act}}$ that is not changed by~$\act$ itself,
i.e., $\var\notin\post{\defi{\act}}$ or $\post{\act}(\var)=\pred{\act}(\var)$,
the integrity constraint in Line~19, which applies in case of $\var\notin\post{\defi{\act}}$,
suppresses a parallel application of actions~$\act'$
such that $\var\in\post{\defi{\act}'}$ and $\post{\act'}(\var)\neq\pred{\act}(\var)$,
while the integrity constraint in Line~13 % of Listing~\ref{lst:baseline}
readily requires~$\var$ to remain unchanged in case $\post{\act}(\var)=\pred{\act}(\var)$.
On the other hand, the situation becomes slightly more involved when
$\var\in\post{\defi{\act}}$ and $\post{\act}(\var)\neq\pred{\act}(\var)$, i.e.,
the application of~$\act$ invalidates its own precondition.
In this case,
no other action~$\act'$ such that $\var\in\post{\defi{\act}'}$ can be applied in parallel,
either because $\post{\act'}(\var)\neq\post{\act}(\var)$ undermines confluence, or since
$\post{\act'}(\var)=\post{\act}(\var)$ disrespects the precondition of~$\act$.
To account for such situations and address all actions invalidating their precondition
regarding~$\var$ at once,
the rule in Line~21 derives an atom over \lstinline{single}/2 to indicate that at most
(and effectively exactly) one action affecting~$\var$ can be applied,
as asserted by the integrity constraint in Line~22.
As a consequence, no action applied in parallel can invalidate the precondition of another action,
so that any serialization leads to the same successor state as obtained in the parallel case.
\begin{lstlisting}[label=lst:forall,caption={Extension of Listing~\ref{lst:baseline} for encoding \astep{} plans},float,firstnumber=17,firstline=3]
:- occurs(A,t), prec(A,X,V), not holds(X,V,t-1).

:- occurs(A,t), prec(A,X,V), not post(A,X,_), not holds(X,V,t).

single(X,t) :- occurs(A,t), prec(A,X,V1), post(A,X,V2), V1 != V2.
:- single(X,t), #count{A : occurs(A,t), post(A,X,V)} > 1.
\end{lstlisting}

\begin{example}\label{ex:forall}
The two sequential plans from Example~\ref{ex:parallel}
correspond to two stable models,
obtained with the encoding of sequential plans given by Listings~\ref{lst:baseline} and~\ref{lst:sequential},
both including the atoms
\lstinline{occurs(}$\act_1$\lstinline{,1)} and
\lstinline{occurs(}$\act_2$\lstinline{,2)}.
In addition, one stable model contains
\lstinline{occurs(}$\act_3$\lstinline{,3)} along with
\lstinline{occurs(}$\act_4$\lstinline{,4)},
and the other
\lstinline{occurs(}$\act_4$\lstinline{,3)} as well as
\lstinline{occurs(}$\act_3$\lstinline{,4)},
thus exchanging the order of applying $\act_3$ and~$\act_4$.
Given that $\act_3$ and $\act_4$ are confluent,
the independence of their application order is expressed by a single stable model,
obtained with the encoding part for \astep\ plans in Listing~\ref{lst:forall}
instead of the one in Listing~\ref{lst:sequential},
comprising
\lstinline{occurs(}$\act_3$\lstinline{,3)} as well as
\lstinline{occurs(}$\act_4$\lstinline{,3)} in addition to
\lstinline{occurs(}$\act_1$\lstinline{,1)} and
\lstinline{occurs(}$\act_2$\lstinline{,2)}.
Note that, even though the set $\{\act_1,\act_2\}$ is confluent,
it is not \astep\ serializable (in~$\init$),
and a parallel application is suppressed in view of the atom
\lstinline{single(}$\var_1$\lstinline{,1)},
derived since $\act_1$ invalidates its precondition regarding~$\var_1$.
Moreover, the requirement that the precondition of an applied action must be established
in the state before permits only
$\langle \{\act_1\},\linebreak[1]\{\act_2\},\linebreak[1]\{\act_3,\act_4\} \rangle$
as \astep\ plan or its corresponding stable model, respectively,
with three sets of actions.
\eofex
\end{example}

Additions to Listing~\ref{lst:baseline} addressing \emph{\estep\ plans} are given in Listing~\ref{lst:exists}.
As before, the integrity constraint in Line~17 is included to assert the precondition of an applied action
to hold in the state referred to by \lstinline{t-1}.
Unlike with \astep\ plans, however,
an applied action may invalidate the precondition of another action,
in which case the other action must come first in a serialization,
and the aim is to make sure that there is some compatible serialization.
To this end, the rule in Lines~19--20 expresses that an action can be safely applied,
as indicated by a respective instance of the head atom \lstinline{apply(A1,t)},
once \emph{all} other actions whose preconditions it invalidates are captured
by corresponding instances of \lstinline{ready(A2,t)}.
The latter provide actions that are not applied or whose application is safe,
i.e., no yet pending action's precondition gets invalidated,
% \comment{JR: This reads a bit complicated, maybe can be improved\ldots}
and are derived by means of the rules in Lines~22 and~23.
In fact, the least fixpoint obtained via the rules in Lines~19--23 covers all actions
if and only if the applied actions do not circularly invalidate their preconditions,
and the integrity constraint in Line~24 prohibits any such circularity,
which in turn means that there is a compatible serialization.
\begin{lstlisting}[label=lst:exists,caption={Extension of Listing~\ref{lst:baseline} for encoding \estep{} plans},float,firstnumber=17,firstline=3]
:- occurs(A,t), prec(A,X,V), not holds(X,V,t-1).

apply(A1,t) :- action(A1),
   ready(A2,t) : post(A1,X,V1), prec(A2,X,V2), A1 != A2, V1 != V2.

ready(A,t) :- action(A), not occurs(A,t).
ready(A,t) :- apply(A,t).
:- action(A), not ready(A,t).
\end{lstlisting}
\begin{lstlisting}[label=lst:acyclic,caption={Replacement of Lines~19--24 in Listing~\ref{lst:exists} by an \lstinline{\#edge} directive},float,firstnumber=19,firstline=5]
#edge((A1,t),(A2,t)) : occurs(A1,t),
                 post(A1,X,V1), prec(A2,X,V2), A1 != A2, V1 != V2.
\end{lstlisting}

Excluding circular interference also lends itself to an alternative implementation
by means of the \lstinline{#edge} directive \cite{gekakaosscwa16a} of \clingo,
in which case built-in acyclicity checking \cite{bogejakasc16a} is used.
A respective replacement of Lines~19--24 is shown in Listing~\ref{lst:acyclic},
where the \lstinline{#edge} directive in Lines~19--20 asserts  edges
from an applied action to all other actions whose preconditions it invalidates,
and acyclicity checking makes sure that the graph induced by applied actions remains acyclic.

\begin{lstlisting}[label=lst:relaxed,caption={Extension of Listing~\ref{lst:baseline} for encoding relaxed \estep{} plans},float,firstnumber=17,firstline=3]
reach(X,V,t) :- holds(X,V,t-1).
reach(X,V,t) :- occurs(A,t), apply(A,t), post(A,X,V).

apply(A1,t) :- action(A1), reach(X,V,t) : prec(A1,X,V);
   ready(A2,t) : post(A1,X,V1), prec(A2,X,V2), A1 != A2, V1 != V2.

ready(A,t) :- action(A), not occurs(A,t).
ready(A,t) :- apply(A,t).
:- action(A), not ready(A,t).
\end{lstlisting}
The encoding part for \emph{relaxed \estep\ plans} in Listing~\ref{lst:relaxed}
deviates from those given so far by not necessitating the
precondition of an applied action to hold in the state before.
Rather, the preconditions of actions applied in parallel may be
established successively,
where confluence along with the condition that an action is
applicable only after other actions whose preconditions it invalidates
have been processed guarantee the existence of a compatible serialization.
In fact, the rules in Lines~20--24 are almost identical to their counterparts in
Listing~\ref{lst:exists}, % and the integrity constraint in Line~27
% readily excludes any circular invalidation of
% the applied actions' preconditions, while 
and   the difference amounts to the additional prerequisite
`\lstinline{reach(X,V,t) : prec(A1,X,V)}'
% for the safe application of an action
in Line~20.
Instances of \lstinline{reach(X,V,t)} are derived by means of the rules in Lines~17 and~18
to indicate fluent values from the state referred to by \lstinline{t-1}
along with postconditions of actions whose application has been determined to be safe.
The prerequisites of the rule in Lines~20--21 thus express that an action can be safely applied
once its precondition is established, possibly by means of other actions preceding it
in a compatible serialization, \emph{and} if it does not invalidate any pending action's precondition.%
\footnote{%
The `\lstinline[basicstyle=\ttfamily\footnotesize]{;}' symbol in Line~20 separates
the (conjunctively connected) conditional literals
`\lstinline[basicstyle=\ttfamily\footnotesize]{reach(X,V,t) :}
 \lstinline[basicstyle=\ttfamily\footnotesize]{ prec(A1,X,V)}'
and
`\lstinline[basicstyle=\ttfamily\footnotesize]{ready(A2,t) : post(A1,X,V1),}
 \lstinline[basicstyle=\ttfamily\footnotesize]{prec(A2,X,V2)},
 \lstinline[basicstyle=\ttfamily\footnotesize]{A1 != A2},
 \lstinline[basicstyle=\ttfamily\footnotesize]{V1 != V2}'.}
Similar to its counterpart in Listing~\ref{lst:exists}, the integrity constraint in Line~25
then makes sure that actions are not applied unless their application is safe in the sense of a
relaxed \estep\ serializable set.

\begin{example}\label{ex:exists}
The \astep\ plan 
$\langle \{\act_1\},\linebreak[1]\{\act_2\},\linebreak[1]\{\act_3,\act_4\} \rangle$
from Example~\ref{ex:parallel}
can be condensed into 
$\langle \{\act_1,\act_2\},\linebreak[1]\{\act_3,\act_4\} \rangle$
when switching to \estep\ serializable sets.
Corresponding stable models obtained with the
encodings given by Listing~\ref{lst:baseline}
along with Listing~\ref{lst:exists} or~\ref{lst:acyclic}
include
\lstinline{occurs(}$\act_1$\lstinline{,1)},
\lstinline{occurs(}$\act_2$\lstinline{,1)},
\lstinline{occurs(}$\act_3$\lstinline{,2)}, and
\lstinline{occurs(}$\act_4$\lstinline{,2)}.
Regarding the \lstinline{#edge} directive in Listing~\ref{lst:acyclic},
these atoms induce the graph
$(\{
   \text{\lstinline{(}}\act_1\text{\lstinline{,1)}},\linebreak[1]
   \text{\lstinline{(}}\act_2\text{\lstinline{,1)}}% ,\linebreak[1]
   % \text{\lstinline{(}}\act_3\text{\lstinline{,2)}},\linebreak[1]
   % \text{\lstinline{(}}\act_4\text{\lstinline{,2)}}
  \},\linebreak[1]
  \{
   \langle\text{\lstinline{(}}\act_2\text{\lstinline{,1)}},\linebreak[1]
          \text{\lstinline{(}}\act_1\text{\lstinline{,1)}}\rangle
  \}
 )$,
which is clearly acyclic.
Its single edge tells us that $\act_1$ must precede $\act_2$ in
a compatible serialization,
while the absence of a cycle means that the application of
$\act_1$ does not invalidate the precondition of~$\act_2$.
In terms of the encoding part in Listing~\ref{lst:exists},
\lstinline{apply(}$\act_1$\lstinline{,1)} and
\lstinline{ready(}$\act_1$\lstinline{,1)} are
derived first,
which in turn allows for deriving
\lstinline{apply(}$\act_2$\lstinline{,1)} and
\lstinline{ready(}$\act_2$\lstinline{,1)}.
The requirement that the precondition of an applied action must be established
in the state before,
which is shared by Listings~\ref{lst:exists} and~\ref{lst:acyclic}, however,
necessitates at least two sets of actions for an \estep\ plan or a corresponding
stable model, respectively.
Unlike that,
the encoding of relaxed \estep\ plans given by Listings~\ref{lst:baseline} and~\ref{lst:relaxed}
yields a stable model
containing
\lstinline{occurs(}$\act_1$\lstinline{,1)},
\lstinline{occurs(}$\act_2$\lstinline{,1)},
\lstinline{occurs(}$\act_3$\lstinline{,1)}, and
\lstinline{occurs(}$\act_4$\lstinline{,1)},
corresponding to the relaxed \estep\ plan 
$\langle \{\act_1,\act_2,\act_3,\act_4\} \rangle$.
The existence of a compatible serialization
is witnessed by first deriving, amongst other atoms,
\lstinline{reach(}$\var_1$\lstinline{,0,1)} and
\lstinline{reach(}$\var_3$\lstinline{,0,1)} in view of~$\init$.
These atoms express that the preconditions of~$\act_1$ and~$\act_2$
are readily established,
so that  
\lstinline{apply(}$\act_1$\lstinline{,1)} along with
\lstinline{reach(}$\var_2$\lstinline{,1,1)}
and
\lstinline{ready(}$\act_1$\lstinline{,1)}
are derived next.
The latter atom indicates that $\act_1$ can be safely applied before~$\act_2$,
which then leads to
\lstinline{apply(}$\act_2$\lstinline{,1)} along with
\lstinline{reach(}$\var_3$\lstinline{,1,1)}.
Together, \lstinline{reach(}$\var_2$\lstinline{,1,1)} and \lstinline{reach(}$\var_3$\lstinline{,1,1)}
     reflect  that the precondition of~$\act_3$ as well as~$\act_4$
can be established by means of~$\act_1$ and~$\act_2$ applied in parallel, so that
\lstinline{apply(}$\act_3$\lstinline{,1)} and
\lstinline{apply(}$\act_4$\lstinline{,1)}
are derived in turn.
\eofex
\end{example}

In order to formalize the soundness and completeness of the presented encodings,
let $\base$ stand for the rule in Line~1 of Listing~\ref{lst:baseline},
$\query{i}$ for the integrity constraint in Line~5 with the parameter~\lstinline{t}
replaced by some integer~$i$, and % likewise
$\step{i}$ for the rules and integrity constraints below the \lstinline{#program} directive in Line~7
with $i$ taken for~\lstinline{t}.
Moreover, we refer to specific encoding parts extending $\step{i}$,
where the parameter~\lstinline{t} is likewise replaced by~$i$, by
$\steps{i}$ for Listing~\ref{lst:sequential},
$\stepa{i}$ for Listing~\ref{lst:forall},
$\stepe{i}$ for Listing~\ref{lst:exists},
$\stepc{i}$ for Line~17 of Listing~\ref{lst:exists} along with Listing~\ref{lst:acyclic}, and 
$\stepr{i}$ for Listing~\ref{lst:relaxed}.
Given that $\stepc{i}$ includes an \lstinline{#edge} directive subject to acyclicity checking,
we understand stable models in the sense of \cite{bogejakasc16a}, i.e.,
the graph induced by a (regular) stable model, which is empty in case of no \lstinline{#edge} directives,
must be acyclic.
\begin{theorem}\label{thm:correct}
Let \fact\ be the set of facts representing a planning task $\langle \vars,\init,\goal,\acts \rangle$,
$\langle \act_1,\dots,\linebreak[1]\act_n \rangle$ be a sequence of actions, and
$\langle \actp_1,\dots,\actp_m \rangle$ be a sequence of sets of actions.
Then, % we have that
\begin{itemize}
\item
$\langle \act_1,\dots,\act_n \rangle$
is a sequential plan
iff
\[\textstyle
 I\cup\base\cup
 \query{0}\cup
 \bigcup_{i=1}^n(\query{i}\cup\step{i}\cup\steps{i})\cup
 \{\text{\lstinline{query(}}n\text{\lstinline{).}}\}
\]
has a stable model~$\smod$ such that
$\{\langle \act, i\rangle \mid 
   \text{\lstinline{occurs(}}\act\text{\lstinline{,}}i\text{\lstinline{)}}\in\smod\}
 =
 \{\langle \act_i,i\rangle \mid 1\leq i\leq n\}$;
\item
$\langle \actp_1,\dots,\actp_m \rangle$
is a \astep\ (resp., \estep\ or relaxed \estep) plan
iff
\[\textstyle
 I\cup\base\cup
 \query{0}\cup
 \bigcup_{i=1}^m(\query{i}\cup\step{i}\cup\stepp{i})\cup
 \{\text{\lstinline{query(}}m\text{\lstinline{).}}\}
\]
with $\stepp{i}=\stepa{i}$ (resp., $\stepp{i}\in\{\stepe{i},\stepc{i}\}$ or $\stepp{i}=\stepr{i}$)
% for $1\leq i\leq m$,
has a stable model~$\smod$ such that
$\{\langle \act, i\rangle \mid 
   \text{\lstinline{occurs(}}\act\text{\lstinline{,}}i\text{\lstinline{)}}\in\smod\}
 =
 \{\langle \act,i\rangle \mid 1\leq i\leq m,\linebreak[1]\act\in\actp_i\}$.
\end{itemize}
\end{theorem}
\begin{proof*} % [Proof (Sketch)]
We distinguish sequential and parallel representations of plans.
\begin{itemize}
\item
Let
\[\textstyle
 P^s(n)
 \ = \
 I\cup\base\cup
 \query{0}\cup
 \bigcup_{i=1}^n(\query{i}\cup\step{i}\cup\steps{i})\cup
 \{\text{\lstinline{query(}}n\text{\lstinline{).}}\}
 \text{.}
\]

($\Rightarrow$)
If $\langle \act_1,\dots,\act_n \rangle$ is a sequential plan,
for any stable model~$\smod$ of $P^s(n)$ such that
$\{\langle \act, i\rangle \mid 
   \text{\lstinline{occurs(}}\act\text{\lstinline{,}}i\text{\lstinline{)}}\in\smod\}
 =
 \{\langle \act_i,i\rangle \mid 1\leq i\leq n\}$,
the rule in $\base$ (Line~1 of Listing~\ref{lst:baseline}) together with
the choice rule in Line~9 of Listing~\ref{lst:baseline},
included in $\step{i}$, and
the integrity constraints in $\step{i}$ (Lines~13 and~15 of Listing~\ref{lst:baseline})
make sure that
\[
  \{
    \langle \var,\val,i\rangle
    \mid 
    \text{\lstinline{holds(}}\var\text{\lstinline{,}}\val\text{\lstinline{,}}i\text{\lstinline{)}}\in \smod
  \}
  =
  \{
    \langle \var,\val,i\rangle
    \mid 
    \var\in\vars, 0\leq i\leq n, \occur{\langle \act_1,\dots,\act_i \rangle}{\init}(\var)=\val
  \}
\text{.}
\]
Hence, $P^s(n)$ cannot have any stable model apart from
\[
\begin{array}{@{}r@{}c@{}r@{}l@{}}
\smod
& \ = \ &
I \cup {} &
\{
  \text{\lstinline{occurs(}}\act_i\text{\lstinline{,}}i\text{\lstinline{)}}
  \mid
  1\leq i\leq n
\}
\\
& & {} \cup {} &
\{
  \text{\lstinline{holds(}}\var\text{\lstinline{,}}\val\text{\lstinline{,}}i\text{\lstinline{)}}
  \mid
  \var\in\vars, 0\leq i\leq n, \occur{\langle \act_1,\dots,\act_i \rangle}{\init}(\var)=\val
\}
\\
& & {} \cup {} &
\{\text{\lstinline{query(}}n\text{\lstinline{)}}\}
\end{array}
\]
such that
$\{\langle \act, i\rangle \mid 
   \text{\lstinline{occurs(}}\act\text{\lstinline{,}}i\text{\lstinline{)}}\in\smod\}
 =
 \{\langle \act_i,i\rangle \mid 1\leq i\leq n\}$.
As one can check, $\smod$ satisfies all rules and integrity constraints in $P^s(n)$,
% where atoms of the form 
% \(
% \text{\lstinline{holds(}}\var\text{\lstinline{,}}\init(\var)\text{\lstinline{,0)}}
% \)
% are derived by the rule in $\base$ % (Line~1 of Listing~\ref{lst:baseline}),
% and the choice rules in $\step{i}$ (Lines~9 and~11 of Listing~\ref{lst:baseline})
% allow for picking the remaining atoms in $\smod\setminus (I\cup\{\text{\lstinline{query(}}n\text{\lstinline{)}}\})$,
so that $\smod$ is indeed a stable model of $P^s(n)$.

($\Leftarrow$)
If $\smod$ is a stable model of $P^s(n)$
such that 
$\{\langle \act, i\rangle \mid 
   \text{\lstinline{occurs(}}\act\text{\lstinline{,}}i\text{\lstinline{)}}\in\smod\}
 =
 \{\langle \act_i,i\rangle \mid 1\leq i\leq n\}$,
the rule in $\base$ establishes that
\[
  \{
    \langle \var,\val\rangle
    \mid 
    \text{\lstinline{holds(}}\var\text{\lstinline{,}}\val\text{\lstinline{,0)}}\in \smod
  \}
  =
  \{
    \langle \var,\val\rangle
    \mid 
    \var\in\vars, \init(\var)=\val
  \}
\text{.}
\]
For any $1\leq i\leq n$, assume that
\[
  \{
    \langle \var,\val\rangle
    \mid 
    \text{\lstinline{holds(}}\var\text{\lstinline{,}}\val\text{\lstinline{,}}i-1\text{\lstinline{)}}\in \smod
  \}
  =
  \{
    \langle \var,\val\rangle
    \mid 
    \var\in\vars, % 0\leq i\leq n, 
    \occur{\langle \act_1,\dots,\act_{i-1} \rangle}{\init}(\var)=\val
  \}
\text{.}
\]
Then, the integrity constraint in Line~17 of Listing~\ref{lst:sequential},
included in $\steps{i}$,
guarantees that 
$\occur{\act_i}{\occur{\langle \act_1,\dots,\act_{i-1} \rangle}{\init}}$
is defined.
Moreover, the choice rule in Line~9 of Listing~\ref{lst:baseline}
and the integrity constraints in $\step{i}$
make sure that
\[
  \{
    \langle \var,\val\rangle
    \mid 
    \text{\lstinline{holds(}}\var\text{\lstinline{,}}\val\text{\lstinline{,}}i\text{\lstinline{)}}\in \smod
  \}
  =
  \{
    \langle \var,\val\rangle
    \mid 
    \var\in\vars, % 0\leq i\leq n, 
    \occur{\langle \act_1,\dots,\act_i \rangle}{\init}(\var)=\val
  \}
\text{.}
\]
Finally, given the fact \lstinline{query(}$n$\lstinline{)},
the integrity constraint in $\query{n}$ (Line~5 of Listing~\ref{lst:baseline})
yields that
$\occur{\langle \act_1,\dots,\act_n \rangle}{\init}(\var)=\goal(\var)$
for each $\var\in\defg$, so that $\langle \act_1,\dots,\act_n \rangle$ is a sequential plan.
\item
Let
\[\textstyle
 P^p(m)
 \ = \
 I\cup\base\cup
 \query{0}\cup
 \bigcup_{i=1}^m(\query{i}\cup\step{i}\cup\stepp{i})\cup
 \{\text{\lstinline{query(}}m\text{\lstinline{).}}\}
\]
where $p\in\{\forall,\exists,E,R\}$.

($\Rightarrow$)
If $\langle \actp_1,\dots,\actp_m \rangle$
is a \astep\ (resp., \estep\ or relaxed \estep) plan, for $1\leq i\leq m$,
let $\stat_i$ be the state such that
$\stat_i(\var)=\post{\act}(\var)$ for each $\act\in\actp_i$ and $\var\in\post{\defi{\act}}$,
and $\stat_i(\var)=\stat_{i-1}(\var)$ for each $\var\in\vars\setminus\bigcup_{\act\in\actp_i}\post{\defi{\act}}$.
Then,
for any stable model~$\smod$ of $P^p(m)$ such that
$\{\langle \act, i\rangle \mid 
   \text{\lstinline{occurs(}}\act\text{\lstinline{,}}i\text{\lstinline{)}}\in\smod\}
 =
 \{\langle \act,i\rangle \mid 1\leq i\leq m,\linebreak[1]\act\in\actp_i\}$,
the rule in $\base$ (Line~1 of Listing~\ref{lst:baseline}) together with
the choice rule in Line~9 of Listing~\ref{lst:baseline},
included in $\step{i}$, and
the integrity constraints in $\step{i}$ (Lines~13 and~15 of Listing~\ref{lst:baseline})
make sure that
\[
  \{
    \langle \var,\val,i\rangle
    \mid 
    \text{\lstinline{holds(}}\var\text{\lstinline{,}}\val\text{\lstinline{,}}i\text{\lstinline{)}}\in \smod
  \}
  =
  \{
    \langle \var,\val,i\rangle
    \mid 
    \var\in\vars, 0\leq i\leq m, \stat_i(\var)=\val
  \}
\text{.}
\]
In the following, we consider the different kinds of plans.
\begin{description}
\item[\astep:]
Assume that $\langle \actp_1,\dots,\actp_m \rangle$ is a \astep\ plan.
Given that an atom of the form
\lstinline{single(}$\var$\lstinline{,}$i$\lstinline{)}
is derived by the rule in $\stepa{i}$ (Line~21 of Listing~\ref{lst:forall}) iff
there is some $\act\in\actp_i$ such that
$\var\in\pred{\defi{\act}}\cap\post{\defi{\act}}$ and $\post{\act}(\var)\neq\pred{\act}(\var)$
for $1\leq i\leq m$,
the program $P^\forall(m)$
cannot have any stable model apart from
\[
\begin{array}{@{}r@{}c@{}r@{}l@{}}
\smod
& \ = \ &
I \cup {} &
\{
  \text{\lstinline{occurs(}}\act\text{\lstinline{,}}i\text{\lstinline{)}}
  \mid
  1\leq i\leq m, \act\in\actp_i
\}
\\
& & {} \cup {} &
\{
  \text{\lstinline{holds(}}\var\text{\lstinline{,}}\val\text{\lstinline{,}}i\text{\lstinline{)}}
  \mid
  \var\in\vars, 0\leq i\leq m, \stat_i(\var)=\val
\}
\\
& & {} \cup {} &
\{
  \text{\lstinline{single(}}\var\text{\lstinline{,}}i\text{\lstinline{)}}
  \mid
  1\leq i\leq m, \act\in\actp_i, \var\in\pred{\defi{\act}}\cap\post{\defi{\act}}, \post{\act}(\var)\neq\pred{\act}(\var)
\}
\\
& & {} \cup {} &
\{\text{\lstinline{query(}}m\text{\lstinline{)}}\}
\end{array}
\]
such that
$\{\langle \act, i\rangle \mid 
   \text{\lstinline{occurs(}}\act\text{\lstinline{,}}i\text{\lstinline{)}}\in\smod\}
 =
 \{\langle \act,i\rangle \mid 1\leq i\leq m,\linebreak[1]\act\in\actp_i\}$.
As for $1\leq i\leq m$,
$\occur{\langle \act_1,\dots,\act_k \rangle}{\linebreak[1]\stat_{i-1}}$
is defined for any sequence $\langle \act_1,\linebreak[1]\dots,\linebreak[1]\act_k \rangle$
such that $\{\act_1,\dots,\act_k\}=\actp_i$, 
we have that $\smod$ satisfies the integrity constraints in $\stepa{i}$ (Lines~17, 19, and~22 of Listing~\ref{lst:forall}),
and one can check that further rules and integrity constraints in $P^\forall(m)$
are satisfied as well, % .
% Moreover, atoms of the form 
% \(
% \text{\lstinline{holds(}}\var\text{\lstinline{,}}\init(\var)\text{\lstinline{,0)}}
% \)
% are derived by the rule in $\base$, % (Line~1 of Listing~\ref{lst:baseline}),
% and the choice rules in $\step{i}$ (Lines~9 and~11 of Listing~\ref{lst:baseline})
% allow for picking the remaining atoms in $\smod\setminus (I\cup\{\text{\lstinline{query(}}n\text{\lstinline{)}}\})$,
so that $\smod$ is indeed a stable model of $P^\forall(m)$.
\item[\estep:]
Assume that $\langle \actp_1,\dots,\actp_m \rangle$ is an \estep\ plan.
Then, for any $1\leq i\leq m$, % and $\actp_i=\{\act_1,\dots,\act_k\}$,
there is some sequence 
$\langle \act_1,\linebreak[1]\dots,\linebreak[1]\act_k \rangle$
such that $\{\act_1,\dots,\act_k\}=\actp_i$ and
$\post{\act_j}(\var)=\pred{\act_{j'}}(\var)$ % =\stat_{i-1}(\var)$
if $\var\in\pred{\defi{\act}_{j'}}\cap\post{\defi{\act}_j}$ for $1\leq j<j'\leq k$.
That is, the edges of the graph
\[
 G_i \ = \
 (
  \acts,
  \{
    \langle\act, \act'\rangle \mid
    \act\in\actp_i, \act'\in\acts\setminus\{\act\}, \var\in\pred{\defi{\act}'}\cap\post{\defi{\act}}, \post{\act}(\var)\neq\pred{\act'}(\var)
  \}
 )
\]
can connect some $\act_{j}$ for $1\leq j\leq k$
to elements of $\acts\setminus\{\act_j,\dots,\act_k\}$ only,
so that $G_i$ is acyclic.
This implies that the graph induced by
\[
\begin{array}{@{}r@{}c@{}r@{}l@{}}
\smod
& \ = \ &
I \cup {} &
\{
  \text{\lstinline{occurs(}}\act\text{\lstinline{,}}i\text{\lstinline{)}}
  \mid
  1\leq i\leq m, \act\in\actp_i
\}
\\
& & {} \cup {} &
\{
  \text{\lstinline{holds(}}\var\text{\lstinline{,}}\val\text{\lstinline{,}}i\text{\lstinline{)}}
  \mid
  \var\in\vars, 0\leq i\leq m, \stat_i(\var)=\val
\}
\\
& & {} \cup {} &
\{\text{\lstinline{query(}}m\text{\lstinline{)}}\}
\end{array}
\]
in view of the \lstinline{#edge} directive in Lines~19--20 of Listing~\ref{lst:acyclic},
which is the disjoint union of graphs $G_i$ for $1\leq i\leq m$,
is acyclic as well.
Moreover, the program $P^E(m)$
cannot have any stable model apart from $\smod$
such that
$\{\langle \act, i\rangle \mid 
   \text{\lstinline{occurs(}}\act\text{\lstinline{,}}i\text{\lstinline{)}}\in\nolinebreak\smod\}
 =
 \{\langle \act,i\rangle \mid 1\leq i\leq m,\linebreak[1]\act\in\actp_i\}$,
and one can check that $\smod$ satisfies all rules and integrity constraints in $P^E(m)$,
so that $\smod$ is indeed a stable model of $P^E(m)$.
Regarding $P^\exists(m)$, 
the acyclicity of $G_i$ yields that
\lstinline{apply(}$\act$\lstinline{,}$i$\lstinline{)}
and
\lstinline{ready(}$\act$\lstinline{,}$i$\lstinline{)}
belong to the least fixpoint of the rules in $\stepe{i}$
(Lines 19--23 of Listing~\ref{lst:exists})
for each $\act\in\acts$ and $1\leq i\leq m$,
and thus $P^\exists(m)$ cannot have any stable model apart from
\[
\begin{array}{@{}r@{}c@{}r@{}l@{}}
\smod'
& \ = \ &
\smod \cup {} &
\{
  \text{\lstinline{apply(}}\act\text{\lstinline{,}}i\text{\lstinline{)}}
  \mid
  \act\in\acts, 1\leq i\leq m
\}
\\
& & {} \cup {} &
\{
  \text{\lstinline{ready(}}\act\text{\lstinline{,}}i\text{\lstinline{)}}
  \mid
  \act\in\acts, 1\leq i\leq m
\}
\end{array}
\]
such that
$\{\langle \act, i\rangle \mid 
   \text{\lstinline{occurs(}}\act\text{\lstinline{,}}i\text{\lstinline{)}}\in\smod'\}
 =
 \{\langle \act,i\rangle \mid 1\leq i\leq m,\linebreak[1]\act\in\actp_i\}$.
As one can check, $\smod'$ satisfies all rules and integrity constraints in $P^\exists(m)$,
so that $\smod'$ is indeed a stable model of $P^\exists(m)$.
%
% for any stable model~$\smod$ of $P^\exists(m)$ or $P^E(m)$ such that
% $\{\langle \act, i\rangle \mid 
%    \text{\lstinline{occurs(}}\act\text{\lstinline{,}}i\text{\lstinline{)}}\in\smod\}
%  =
%  \{\langle \act,i\rangle \mid 1\leq i\leq m,\linebreak[1]\act\in\actp_i\}$,
% the integ
%
\item[relaxed \estep:]
Assume that $\langle \actp_1,\dots,\actp_m \rangle$ is a relaxed \estep\ plan.
Then, for any $1\leq i\leq m$, % and $\actp_i=\{\act_1,\dots,\act_k\}$,
there is some sequence 
$\langle \act_1,\linebreak[1]\dots,\linebreak[1]\act_k \rangle$
such that $\{\act_1,\dots,\act_k\}=\actp_i$ and
$\occur{\act_j}{\occur{\langle \act_1,\dots,\act_{j-1} \rangle}{\stat_{i-1}}}$
is defined for $1\leq j\leq k$,
which implies that
$\post{\act_j}(\var)=\pred{\act_{j'}}(\var)$ % =\stat_{i-1}(\var)$
if $\var\in\pred{\defi{\act}_{j'}}\cap\post{\defi{\act}_j}$ for $j<j'\leq k$.
Given this, the least fixpoint of the rules in $\stepr{i}$
(Lines 17--24 of Listing~\ref{lst:relaxed}) contains
\lstinline{reach(}$\var$\lstinline{,}$\val$\lstinline{,}$i$\lstinline{)}
for each $\var\in\vars$ and $\val\in\{\stat_{i-1}(\var),\stat_i(\var)\}$,
\lstinline{apply(}$\act$\lstinline{,}$i$\lstinline{)}
for each $\act\in\acts$ such that 
$\pred{\act}(\var)\in\{\stat_{i-1}(\var),\stat_i(\var)\}$
for all $\var\in\pred{\defi{\act}}$, as well as
\lstinline{ready(}$\act$\lstinline{,}$i$\lstinline{)}
for each $\act\in\acts$,
and thus $P^R(m)$ cannot have any stable model apart from 
\[
\begin{array}{@{}r@{}c@{}r@{}l@{}}
\smod
& \ = \ &
I \cup {} &
\{
  \text{\lstinline{occurs(}}\act\text{\lstinline{,}}i\text{\lstinline{)}}
  \mid
  1\leq i\leq m, \act\in\actp_i
\}
\\
& & {} \cup {} &
\{
  \text{\lstinline{holds(}}\var\text{\lstinline{,}}\val\text{\lstinline{,}}i\text{\lstinline{)}}
  \mid
  \var\in\vars, 0\leq i\leq m, \stat_i(\var)=\val
\}
\\
& & {} \cup {} &
\{
  \text{\lstinline{reach(}}\var\text{\lstinline{,}}\val\text{\lstinline{,}}i\text{\lstinline{)}}
  \mid
  \var\in\vars, 1\leq i\leq m, \val\in\{\stat_{i-1}(\var),\stat_i(\var)\}
\}
\\
& & {} \cup {} &
\{
  \text{\lstinline{apply(}}\act\text{\lstinline{,}}i\text{\lstinline{)}}
  \mid
  \act\in\acts, 1\leq i\leq m,
  \pred{\act}(\var)\in\{\stat_{i-1}(\var),\stat_i(\var)\} \text{ for all } \var\in\pred{\defi{\act}}
\}
\\
& & {} \cup {} &
\{
  \text{\lstinline{ready(}}\act\text{\lstinline{,}}i\text{\lstinline{)}}
  \mid
  \act\in\acts, 1\leq i\leq m
\}
\\
& & {} \cup {} &
\{\text{\lstinline{query(}}m\text{\lstinline{)}}\}
\end{array}
\]
such that
$\{\langle \act, i\rangle \mid 
   \text{\lstinline{occurs(}}\act\text{\lstinline{,}}i\text{\lstinline{)}}\in\smod\}
 =
 \{\langle \act,i\rangle \mid 1\leq i\leq m,\linebreak[1]\act\in\actp_i\}$.
As one can check, $\smod$ satisfies all rules and integrity constraints in $P^R(m)$,
so that $\smod$ is indeed a stable model of $P^R(m)$.
\end{description}

($\Leftarrow$)
If $\smod$ is a stable model of $P^p(m)$ such that
$\{\langle \act, i\rangle \mid 
   \text{\lstinline{occurs(}}\act\text{\lstinline{,}}i\text{\lstinline{)}}\in\smod\}
 =
 \{\langle \act,i\rangle \mid 1\leq i\leq m,\linebreak[1]\act\in\actp_i\}$,
for any $1\leq i\leq m$,
the choice rule in Line~9 of Listing~\ref{lst:baseline}
makes sure that $\stat_i$ such that
$\stat_i(\var)=\val$ iff 
\lstinline{holds(}$\var$\lstinline{,}$\val$\lstinline{,}$i$\lstinline{)}${}\in\smod$
is a state.
Moreover, the rule in $\base$ together with
the integrity constraints in $\step{i}$
yield that
$\stat_i(\var)=\post{\act}(\var)$ for each $\act\in\actp_i$ and $\var\in\post{\defi{\act}}$,
which implies that $\actp_i$ is confluent,
and $\stat_i(\var)=\stat_{i-1}(\var)$ for each $\var\in\vars\setminus\bigcup_{\act\in\actp_i}\post{\defi{\act}}$.
In view of the fact \lstinline{query(}$m$\lstinline{)},
the integrity constraint in $\query{m}$ (Line~5 of Listing~\ref{lst:baseline})
further asserts that $\stat_m(\var)=\goal(\var)$ for each $\var\in\defg$.
It remains to show that $\actp_i=\{\act_1,\dots,\act_k\}$
is \astep\ (resp., \estep\ or relaxed \estep) serializable in $\stat_{i-1}$
for $p=\forall$ (resp., $p\in\{\exists,E\}$ or $p=R$).
\begin{description}
\item[$p=\forall$:] 
Let 
$\langle \act_1,\linebreak[1]\dots,\linebreak[1]\act_k \rangle$
be some sequence such that $\{\act_1,\dots,\act_k\}=\actp_i$.
Then, the integrity constraint in Line~17 of Listing~\ref{lst:forall} 
% included in $\stepa{i}$,
guarantees that $\occur{\act_j}{\stat_{i-1}}$ is defined for $1\leq j\leq k$,
and assume that $\occur{\langle \act_1,\dots,\act_{j-1} \rangle}{\stat_{i-1}}$
is defined as well.
For any $\var\in\pred{\defi{\act}_j}\cap\bigcup_{1\leq j'<j}\post{\defi{\act}_{j'}}$,
we have that $\occur{\langle \act_1,\dots,\act_{j-1} \rangle}{\stat_{i-1}}(\var)=\stat_i(\var)$.
If $\var\notin\post{\defi{\act}_j}$, the integrity constraint in Line~19 of Listing~\ref{lst:forall}
yields that $\stat_i(\var)=\pred{\act_j}(\var)=\stat_{i-1}(\var)$.
Otherwise, if $\var\in\post{\defi{\act}_j}$,
the integrity constraint in Line~22 of Listing~\ref{lst:forall} implies that
\lstinline{single(}$\var$\lstinline{,}$i$\lstinline{)}${}\notin\smod$,
which means that the prerequisites in the instance
\begin{lstlisting}[numbers=none,frame=none,escapechar=?]
single(?$\var$?,?$i$?) :- occurs(?$\act_j$?,?$i$?),?\!? prec(?$\act_j$?,?$\var$?,?$v_1$?),?\!? post(?$\act_j$?,?$\var$?,?$v_2$?),?\!? ?$v_1$?!=?\;$v_2$?.
\end{lstlisting}
of the rule in Line~21 of Listing~\ref{lst:forall} cannot hold and
$\stat_i(\var)=\post{\act_j}(\var)=\pred{\act_j}(\var)=\stat_{i-1}(\var)$ must be the case.
This shows that 
$\occur{\langle \act_1,\dots,\act_{j-1} \rangle}{\stat_{i-1}}(\var)=\pred{\act_j}(\var)$, % =\stat_i(\var)=\stat_{i-1}(\var)=\pred{\act_j}(\var)$,
% from which we conclude 
so that 
$\occur{\langle \act_1,\dots,\act_j \rangle}{\stat_{i-1}}$  and, in particular, 
$\occur{\langle \act_1,\dots,\act_k \rangle}{\stat_{i-1}}$ is defined.
\item[$p\in\{\exists,E\}$:]
For any $\act\in\actp_i$,
the integrity constraint in Line~17 of Listing~\ref{lst:exists}
guarantees that $\occur{\act}{\stat_{i-1}}$ is defined.
Regarding $P^E(m)$,
the \lstinline{#edge} directive in Lines~19--20 of Listing~\ref{lst:acyclic} % ,
% included in $\stepc{i}$,
makes sure that the graph
\[
 G_i \ = \
 (
  \acts,
  \{
    \langle\act, \act'\rangle \mid
    \act\in\actp_i, \act'\in\acts\setminus\{\act\}, \var\in\pred{\defi{\act}'}\cap\post{\defi{\act}}, \post{\act}(\var)\neq\pred{\act'}(\var)
  \}
 )
\]
is acyclic.
Hence, there is some sequence
$\langle \act_1,\linebreak[1]\dots,\linebreak[1]\act_k \rangle$
such that $\{\act_1,\dots,\act_k\}=\actp_i$ and no edge connects 
any element of $\{\act_1,\dots,\act_{j-1}\}$ to $\act_j$ in $G_i$ for $1\leq j\leq k$.
This yields that $\post{\act_{j'}}(\var)=\pred{\act_j}(\var)=\stat_{i-1}(\var)$
if $\var\in\pred{\defi{\act}_j}\cap\post{\defi{\act}_{j'}}$ for $1\leq j'<j$,
so that 
$\occur{\langle \act_1,\dots,\act_j \rangle}{\stat_{i-1}}$ and, in particular, 
$\occur{\langle \act_1,\dots,\act_k \rangle}{\stat_{i-1}}$ is defined.
Concerning $P^\exists(m)$,
the integrity constraint in Line~24 of Listing~\ref{lst:exists} implies that
\lstinline{ready(}$\act$\lstinline{,}$i$\lstinline{)}${}\in\smod$
for each $\act\in\acts$.
The rules in Lines~22 and~23 of Listing~\ref{lst:exists} further yield that
\lstinline{apply(}$\act$\lstinline{,}$i$\lstinline{)}${}\in\smod$
for each $\act\in\actp_i$.
Given the rule in Lines 19--20 of Listing~\ref{lst:exists}, for any $1\leq j\leq k$,
there must be some $\act_j\in\actp_i$ such that
$\{\act_{j'}\in\actp_i\mid j<j'\leq k,\linebreak[1] \var\in\pred{\defi{\act}_{j'}}\cap\post{\defi{\act}_j},\linebreak[1] \post{\act_j}(\var)\neq\pred{\act_{j'}}(\var)\}=\emptyset$.
In turn, we have that $\post{\act_{j'}}(\var)=\pred{\act_j}(\var)=\stat_{i-1}(\var)$
if $\var\in\pred{\defi{\act}_j}\cap\post{\defi{\act}_{j'}}$ for $1\leq j'<j$,
so that 
$\occur{\langle \act_1,\dots,\act_j \rangle}{\stat_{i-1}}$ and, in particular, 
$\occur{\langle \act_1,\dots,\act_k \rangle}{\stat_{i-1}}$ is defined.
\item[$p=R$:]
In view of the rule in Line~17 of Listing~\ref{lst:relaxed},
we have that
\lstinline{reach(}$\var$\lstinline{,}$\val$\lstinline{,}$i$\lstinline{)}${}\in\smod$
if $\stat_{i-1}(\var)=\val$.
The integrity constraint in Line~25 of Listing~\ref{lst:relaxed} further implies that
\lstinline{ready(}$\act$\lstinline{,}$i$\lstinline{)}${}\in\smod$
for each $\act\in\acts$,
and the rules in Lines~23 and~24 of Listing~\ref{lst:relaxed} yield that
\lstinline{apply(}$\act$\lstinline{,}$i$\lstinline{)}${}\in\smod$
for each $\act\in\actp_i$.
Given the rules in Lines~18 and 20--21 of Listing~\ref{lst:relaxed},
for any $1\leq j\leq k$,
there must be some $\act_j\in\actp_i$ such that
$\{\act_{j'}\in\nolinebreak\actp_i\mid\linebreak[1] j<\nolinebreak j'\leq k,\linebreak[1] \var\in\pred{\defi{\act}_{j'}}\cap\post{\defi{\act}_j},\linebreak[1] \post{\act_j}(\var)\neq\pred{\act_{j'}}(\var)\}=\emptyset$
and
$\pred{\act_j}(\var)\in\{\stat_{i-1}(\var)\}\cup
 \{\post{\act_{j'}}(\var)\mid\linebreak[1] 1\leq\nolinebreak j'<\nolinebreak j,\linebreak[1] \var\in\post{\defi{\act}_{j'}}\}$
for all $\var\in\pred{\defi{\act}_j}$.
Hence, for each $\var\in\pred{\defi{\act}_j}$, we have that $\pred{\act_j}(\var)=\stat_{i-1}(\var)$ 
and $\var\notin\bigcup_{1\leq j'< j}\post{\defi{\act}_{j'}}$ or
$\post{\act_{j'}}(\var)=\pred{\act_j}(\var)$
if $\var\in\post{\defi{\act}_{j'}}$ for $1\leq j'<j$,
so that 
$\occur{\langle \act_1,\dots,\act_j \rangle}{\stat_{i-1}}$ and, in particular, 
$\occur{\langle \act_1,\dots,\act_k \rangle}{\stat_{i-1}}$ is defined.
\eofex
\end{description}
\end{itemize}
\end{proof*}
%
%%% Local Variables:
%%% mode: latex
%%% TeX-master: "paper"
%%% End:
%
Let us note that, with each of the considered encodings,
any plan corresponds to a unique stable model,
as the latter is fully determined by atoms over \lstinline{occurs}/2, i.e.,
corresponding (successor) states as well as auxiliary predicates
functionally depend on the applied actions.
Regarding the encoding part for relaxed \estep\ plans in Listing~\ref{lst:relaxed},
we mention that acyclicity checking cannot (in an obvious way) be used
instead of rules dealing with the safe application of actions.
To see this, consider $\langle \vars,\init,\goal,\acts \rangle$ with
$\vars=\{\var_1,\var_2,\var_3\}$ such that $\dom{\var_1}=\dom{\var_2}=\dom{\var_3}=\{0,1\}$,
$\init=\{\var_1=0,\linebreak[1]\var_2=0,\linebreak[1]\var_3=0\}$,
$\goal=\{\var_3=1\}$, and
$\acts=\{\act_1,\act_2\}$,
where
$\act_1=\langle \emptyset,\linebreak[1]\{\var_1=1,\linebreak[1]\var_2=1\} \rangle$ and
$\act_2=\langle \{\var_1=1,\linebreak[1]\var_2=0\},\linebreak[1]\{\var_3=1\} \rangle$.
There is no sequential plan for this task since only $\act_1$ is applicable in $\init$,
but its application invalidates the precondition of~$\act_2$.
Concerning the (confluent) set $\{\act_1,\act_2\}$, 
the graph 
$(\{
   \text{\lstinline{(}}\act_1\text{\lstinline{,1)}},\linebreak[1]
   \text{\lstinline{(}}\act_2\text{\lstinline{,1)}}
  \},\linebreak[1]
  \{
   \langle\text{\lstinline{(}}\act_1\text{\lstinline{,1)}},\linebreak[1]
          \text{\lstinline{(}}\act_2\text{\lstinline{,1)}}\rangle
  \}
 )$
is acyclic and actually includes the information that $\act_2$ should precede $\act_1$
in any compatible serialization.
However, if the prerequisite in Line~21 of Listing~\ref{lst:relaxed} were dropped
to ``simplify'' the encompassing rule, the application of $\act_1$ would be
regarded as safe, and then the precondition of $\act_2$ would seem established as well.
That is, it would be unsound to check the noncircularity of establishment and invalidation of preconditions
in separation, no matter the respective implementation techniques.

As regards encoding techniques, common ASP-based approaches, e.g., \cite{lifschitz02a},
define successor states, i.e., the predicate \lstinline{holds}/3, in terms of actions
given by atoms over \lstinline{occurs}/2.
Listing~\ref{lst:baseline}, however, includes a respective choice rule, which puts it
inline with SAT planning, where our intention is to avoid asymmetries between fluents
and actions, as either of them would in principle be sufficient to indicate plans \cite{kamcse96a}.
Concerning (relaxed) \estep\ plans,
the encoding parts in Listings~\ref{lst:exists} and~\ref{lst:relaxed}
make use of the built-in well-foundedness requirement in ASP and do, unlike \cite{riheni06a},
not unfold the order of actions applied in parallel.
In contrast to the SAT approach to relaxed \estep\ plans in \cite{wehrin07a},
we do not rely on a fixed (static) order of actions,
and to our knowledge, no encoding similar to the one in Listing~\ref{lst:relaxed}
has been proposed so far.

%%% Local Variables:
%%% mode: latex
%%% TeX-master: "paper"
%%% End:

\section{A Multishot ASP Planner}
\label{sec:planner}
Planning encodings must be used with a strategy for fixing the plan length. 
For example, the first approaches to planning in SAT and ASP % \cite{kausel92b,dineko97a}
follow a sequential algorithm starting from 0 and successively incrementing the length by~1
until a plan is found.
% This strategy guarantees shortest plans.

% ------------------------------------------------------------
\begin{figure}
\includegraphics{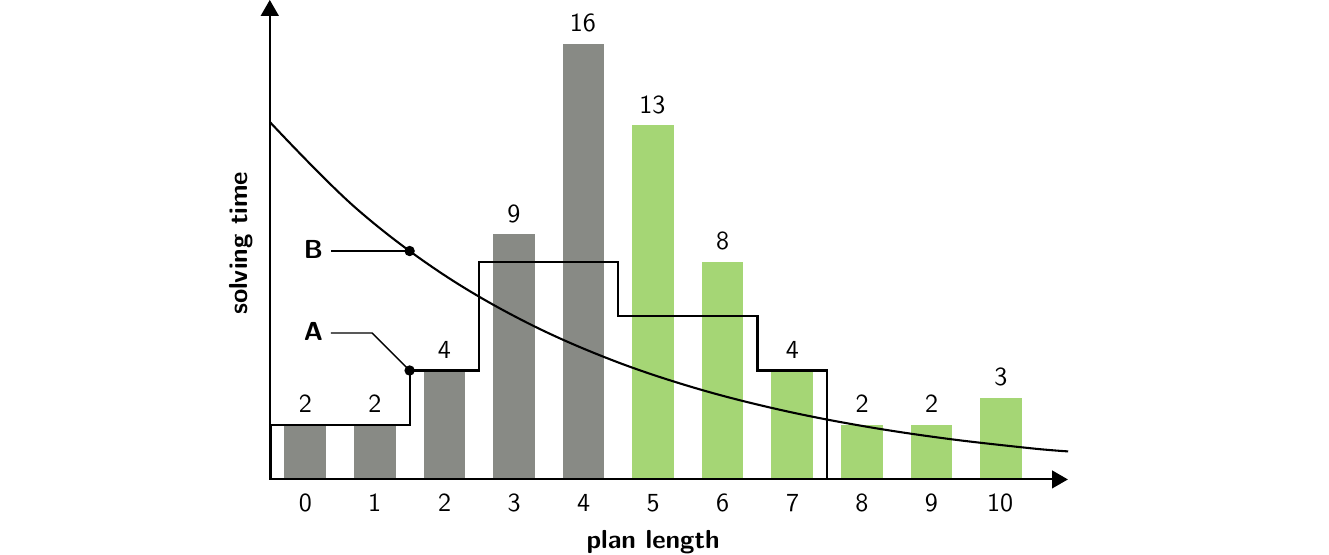}
\caption{Exemplary solving times required by the planning algorithms \alga{}{} and \algb{}{}}
\label{planner:fig}
\end{figure}
% ------------------------------------------------------------
%
For parallel planning in SAT, more flexible strategies were proposed in \cite{riheni06a}, based on the following ideas.
First, minimal parallel plans do not coincide with shortest sequential plans.
Hence, it is unclear whether parallel plans should be minimal.
Second, solving times for different plan lengths follow a certain pattern,
which can be exploited.
To illustrate this, consider the solving times of a typical instance in Figure \ref{planner:fig}.
For lengths 0 to 4, in gray, the instance is unsatisfiable, and 
time grows exponentially.
Then, the first satisfiable instances, in green, are still hard,
but they become easier for greater plan lengths.
However, for even greater plan lengths, the solving time increases again because the size becomes larger.
Accordingly,
\cite{riheni06a} suggests not to minimize parallel plan length,
but rather make use of \emph{planning algorithms} that 
avoid costly unsatisfiable parts by  moving early to easier satisfiable lengths.

The sequential algorithm (\algs{}{}) solves the instance in Figure \ref{planner:fig} in $46$ time units,
% (say seconds), 
viz.\ $2+2+4+9+16+13$, 
by trying plan lengths 0 to 4 until it finds a plan at 5.
The idea of algorithm \alga{}{} \cite{riheni06a} is to simultaneously consider $n$ plan lengths. 
In our example, fixing $n$ to $5$, \alga{}{} starts with lengths 0 to 4.
After $2$ time units, lengths 0 and 1 are finished, and 5 and 6 are added.
Another $2$ units later, length 2 is finished, and 7 is started.
Finally, after $4$ more units per length from 3 to 7, length 7 yields a plan.
The times spent by \alga{}{} for each length are indicated in Figure \ref{planner:fig},
and summing them up % the times spend for all of them % individual lengths yields
amounts to $40$ time units in total.
%
%Observe how in this case the easy instance could be solved before terminating the previous costly lengths.
%
%In algorithm $B$ there is one process that simulates and infinite number of processes, 
%every one having assigned less time than the previous ones.
%
Algorithm \algb{}{} \cite{riheni06a} distributes time nonuniformly over plan lengths:
if length $n$ is run for $t$ time units,
then lengths $n+i$ are run for $t*\gamma^i$ units,
where $i\geq 1$ and $\gamma$ lies between $0$ and $1$.
In our example, we set $\gamma$ to $0.8$, % . % $\frac{4}{5}$.
and the amount~$t$ of time spent on the initially shortest length~0 is thus
multiplied by $0.8^i$ for lengths $i\geq 1$.
Note that, in practice, only lengths whose assigned time is above some threshold are
indeed run, which restricts the plan lengths to consider simultaneously.
While searching for a plan, the shortest unfinished length~$n$
and its assigned time~$t$ successively increase,
so that $t*\gamma^i$ grows beyond the threshold of running for greater plan lengths $n+i$. 
%
% \new{
% Initially, length $0$ is assigned some amount of time.
% %
% During the execution of the algorithm, 
% the time $t$ assigned to the shortest unfinished length $n$ is successively incremented,
% and is used to determine the times $t*\gamma^i$ of the longer lengths $n+i$.
% %
% In practice, only the lengths whose assigned time is above some threshold are run.
% }
%
Regarding our example with $\gamma=0.8$,
when length 3 has been run for $6$ time units, 
previous lengths are already finished, and
the times for the following lengths are given by curve \algb{}{} in Figure \ref{planner:fig}.
At this point, length 8 is assigned $2$ units ($\lceil 6*0.8^5 \rceil$) % ($\lceil 6*\frac{4}{5}^5 \rceil$)
and yields a plan, 
leading to a total time of $38$ units:
$8$ units for lengths 0 to 2, and $30$ for the rest.
(The $30$ units correspond to the area under the curve from length 3 on.)
Note that both \alga{}{} and \algb{}{} find a plan before finishing the hardest instances
and, in practice,
% \alga{}\ and \algb{}\
often save significant time over \algs{}{}.

We adopted algorithms \alga{}{} (yielding \algs{}{} when $n$ is set to~$1$)
%\comment{P: This is confusing, as this can be interpreted such that \alga{}{} is the same as \algs{}{} with $n = 1$ and not the other way around.}
and \algb{}{}, and implemented them as planning strategies of \plasp\ via multishot ASP solving.
In general,
they can be applied to any incremental encoding complying with the threefold structure
of \lstinline{base}, \lstinline{step(t)}, and \lstinline{check(t)} subprograms.
Assuming that the subprograms adhere to \clingo's modularity condition \cite{gekakasc14b},
they are assembled to ASP programs of the form
\[\textstyle
P(n)
\ = \
\text{\lstinline{base}} \cup\bigcup_{i=0}^n \text{\lstinline{check(}}i\text{\lstinline{)}} \cup \bigcup_{i=1}^n \text{\lstinline{step(}}i\text{\lstinline{)}}  
\]
where $n$ gives the length of the unrolled encoding. %, 
%$m \leq n$ is the time point at which the query atom holds.
%
%Observe that with $m<n$ the query atom can be used to require the goal 
%\emph{before} the end of the plan. 
%
The planner then looks for an integer $n$ such that $P(n) \cup \queryatom{n}$ is satisfiable, 
and algorithms \algs{}{}, \alga{}{}, and \algb{}{}
provide different strategies to approach such an integer.
% In practice, for solving a planning problem, 
% one must define appropriate subprograms \lstinline{base}, \lstinline{step} and \lstinline{check}, 
% so that the solutions returned by the planner correspond to valid plans.

The planner%
\footnote{\url{https://github.com/potassco/planner}}
is implemented using \clingo's multishot solving capacities,
where a   \clingo\ object grounds and solves incrementally. 
This approach avoids extra grounding efforts  and 
allows for taking advantage of previously learned constraints. %  information.
The planner simulates the parallel processing of different plan lengths
by interleaving sequential subtasks. % executions.
%
% In the example of Algorithm A, 
% the solver starts running sequentially lengths $0$ to $4$ for a short time. %
% %
% %\footnote{Concretely, every length runs for $100$ restarts.}
% %
% The sequence is run again and again, 
% until length $0$ finishes and length $5$ is added.
% %
% Afterwards, it is the turn of lengths $1$ to $5$, and the process continues.
%
To this end,
the \clingo\ object is used to successively unroll an incremental encoding up to integer(s)~$n$.
% 
% When a goal at a larger step $m>n$ must be checked,
% the encoding is further unrolled until $m$.
%
In order to solve a subtask for some $m<n$,
the unrolled part $P(n)$ is kept intact, while
\lstinline{query(}$m$\lstinline{)} is set to true instead of
\lstinline{query(}$n$\lstinline{)}.
% the query atom is set true at time point $m$ instead of $n$.
%
That is, the search component of \clingo\ has to establish conditions in \lstinline{check(}$m$\lstinline{)},
even though the encoding is unrolled up to $n \geq m$.
%
% IMPROVE NEXT LINES
%
For this approach to work, we require that
$P(m) \cup \queryatom{m}$ is satisfiable if and only if $P(n) \cup \queryatom{m}$ is satisfiable
for $0 \leq m \leq n$.
%
% This guarantees that a solution of length $n$ also works for length $m$, 
% and allows for retrieving it for length $m$.
%
An easy way to guarantee this property is
to tolerate idle states in between~$m$ and~$n$,
as is the case with the encodings given in Section~\ref{sec:encodings}.
% the nonexecution of actions between $m$ and $n$, 
% so that no changes between $m$ and $n$ occur.

While planning algorithms tackle the issue of finding a sufficient plan length,
the choice of an underlying planning encoding remains, i.e., whether to take
$\step{i}\cup\steps{i}$,
$\step{i}\cup\stepa{i}$,
$\step{i}\cup\stepe{i}$,
$\step{i}\cup\stepc{i}$, or
$\step{i}\cup\stepr{i}$
according to the terminology used in Theorem~\ref{thm:correct}
for the subprogram \lstinline{step(}$i$\lstinline{)} of $P(n)$.
On the one hand, the encoding $\step{i}\cup\stepr{i}$ of relaxed \estep\ plans
is guaranteed to become satisfiable first,
where the minimal parallel plan length may still coincide with the sequential encoding
$\step{i}\cup\steps{i}$ in the ``worst'' case of an inherently sequential planning task.
On the other hand,
parallel encodings introduce overhead for checking the existence of a compatible
serialization.
This particularly applies to $\stepe{i}$, $\stepc{i}$, and $\stepr{i}$, aiming at
(relaxed) \estep\ plans,
as the conditions they encode refer to pairs of actions whose
preconditions and postconditions interfere.
Such quadratic behavior is problematic for planning tasks involving a large number of actions,
and our experiments in Section~\ref{sec:experiments} indeed incorporate domains
where instances yield several thousand actions.
As a consequence, it is sometimes desirable to keep the efforts of checking
whether a set of actions is serializable low.
Moreover, investigations of common benchmark domains for planning systems \cite{riheni06a,rintanen12a}
showed that circular interference is in many cases impossible,
so that some serialization will exist for any (confluent) set of actions.
This observation along with the aforementioned efficiency considerations regarding the (ground) representation
of serialization conditions motivate us to augment the planner with \emph{guess-and-check}
facilities, detailed in the following.

The general idea of the guess-and-check approach \cite{eitpol06a}
is to encode a problem by a pair $\langle G, C \rangle$ of programs,
where a stable model~$\smod$ of~$G$ constitutes a solution if $C\cup\smod$ is unsatisfiable.
In the context of ASP planning,
the role of~$G$ is to generate stable models providing sequences of sets of actions,
and~$C$ checks whether some serialization yields a sequential plan.
For making ``educated'' guesses,
the program~$G$ we propose combines facts representing a planning task
with the incremental encoding in Listing~\ref{lst:baseline}
and the integrity constraint in Line~17,
shared by Listings \ref{lst:sequential}--\ref{lst:exists},
for asserting the preconditions of applied actions.
As a consequence, a stable model~$\smod$ of~$G$ is such that 
all preconditions and postconditions hold for a set of actions to be applied in parallel,
which also makes sure that the set is confluent,
while (the absence of) circular interference remains to be checked.
The latter can be accomplished by taking the
atoms over %\lstinline{holds}/3 and 
\lstinline{occurs}/2 from $\smod$ as facts
together with a program~$C$ comprising the fact representation of a planning task,
the rules in Lines 19--23 of Listing~\ref{lst:exists},
and an encoding part as follows:%
\begin{lstlisting}[numbers=none]
#program step(t).

cycle(t) :- action(A), not ready(A,t).

#program check(t).

:- query(t), not cycle(T) : T = 1..t.
\end{lstlisting}
Note that instances of the rule in the \lstinline{step(t)} subprogram yield \lstinline{cycle(t)}
if applied actions are pending, i.e., the respective instances of \lstinline{ready(A,t)}
remain underivable,
as each such action invalidates another pending action's precondition.
In fact,
an atom \lstinline{cycle(}$i$\lstinline{)} means that the set of actions~$\act$
such that \lstinline{ready(}$\act$\lstinline{,}$i$\lstinline{)} does not hold
is not (relaxed) \estep\ serializable in any state.
Given this,
the integrity constraint in \lstinline{check(t)} requires that
some set of actions is not \estep\ serializable in view of a circular
invalidation of preconditions.
Letting $m=\max\{i \mid \text{\lstinline{occurs(}}\act\text{\lstinline{,}}i\text{\lstinline{)}}\in\smod\}$,
a stable model of
$C
% \cup \linebreak[1]
% \{\text{\lstinline{holds(}}\var\text{\lstinline{,}}\val\text{\lstinline{,}}i\text{\lstinline{).}} \mid \linebreak[1]
%   \text{\lstinline{holds(}}\var\text{\lstinline{,}}\val\text{\lstinline{,}}i\text{\lstinline{)}}\in\smod\}
 \cup \linebreak[1]
 \{\text{\lstinline{occurs(}}\act\text{\lstinline{,}}i\text{\lstinline{).}} \mid \linebreak[1]
   \text{\lstinline{occurs(}}\act\text{\lstinline{,}}i\text{\lstinline{)}}\in\smod\}
 \cup \linebreak[1]
 \{\text{\lstinline{query(}}m\text{\lstinline{).}}\}$
thus tells us that the sequence of sets of actions from~$\smod$ is not (relaxed) \estep\ serializable,
while the absence of stable models indicates the existence of a compatible serialization.

In case the program~$C$ together with facts for some sequence $\langle \actp_1,\dots,\actp_m \rangle$
of actions from a stable model~$\smod$ of~$G$ is satisfiable,
the corresponding stable model~$\smod'$ is such that
$\actp_i'=\{\act\in\actp_i \mid \text{\lstinline{ready(}$\act$\lstinline{,}$i$\lstinline{)}}\notin M'\}$
is nonempty for at least one $1\leq i\leq m$.
Given that each nonempty $\actp_i'$ is not (relaxed) \estep\ serializable,
a guess-and-check control loop could augment $G$ with constraints suppressing
a parallel application of the actions in $\actp_i'$ to eliminate~$\smod$,
but not any \estep\ plan, and search for another stable model instead.
In fact, we have tried several options to utilize the information from
a stable (counter-)model~$\smod'$, i.e., extending $G$ with constraints
that deny a parallel application of $\actp_i'$ and supersets thereof at
the $i$th or all positions of a sequence of actions, respectively,
in order to generate alternative sequences.
However, we found that the domains used for our experiments in Section~\ref{sec:experiments}
belong to two rather extreme categories:
either the sequence of actions generated first directly yields a compatible serialization,
given that the available actions cannot interfere,
or a vast number of sequences that are not \estep\ serializable is successively generated,
so that denying the parallel application of particular sets of actions turns out to be ineffective.
Hence, switching from the program~$G$ given above to one of the parallel encodings
provided in Section~\ref{sec:encodings}, in case the sequence of sets of actions generated first
is not \estep\ serializable, constitutes a better option to implement the guess-and-check approach.
In Section~\ref{sec:experiments}, we particularly investigate the strategy to switch to
the encoding of \astep\ plans in Listing~\ref{lst:forall},
as it avoids the aforementioned issue of referring to pairs of actions
to express serialization conditions.
Technically, the switch from $G$ to the encoding of \astep\ plans is
accomplished by including Lines~19--22 of Listing~\ref{lst:forall}
in a separate subprogram instead of \lstinline{step(t)},
which is then instantiated for the same integers starting from~\lstinline{1}
as used for \lstinline{step(t)} in case the first sequence of sets of actions obtained with~$G$
happens to be not \estep\ serializable.
While the guess-and-check strategies discussed here aim at an efficient problem representation
by skipping serialization conditions unless they are needed,
we note that other applications, e.g., in conformant or temporal planning
\cite{cipitr08a,fisher08a}, may also harness the approach to perform more sophisticated checks.

Our planner is further equipped with a planning-specific \emph{heuristic},
%introduced in \cite{gekaotroscwa13a} and,
inspired by \cite{rintanen12a} and devised in \cite{gekaotroscwa13a} 
within a framework for domain-specific heuristics in ASP.
The general idea is to extend the search heuristic of \clingo\ by
associating each atom with a \emph{level} ($0$ by default)
and a \emph{sign}, which can be undefined (by default), true, or false.
During solving, these values get modified by the activation of 
\lstinline{#heuristic} directives, and
the search heuristic of \clingo\ then selects an atom at the highest level
and sets it to true or false according to its associated sign,
switching to the default sign heuristic if that sign is undefined.
The domain-specific heuristic for planning,
which aims at propagating fluent values backwards in time,
starting from those in the goal of a planning task,
is specified in terms of \lstinline{#heuristic} directives as follows:
%
% Corresponding \lstinline{#heuristic} directives are as follows:
% The heuristic is declaratively specified by the following \lstinline{#heuristic} directives of \clingo:
%
%\new
\begin{lstlisting}[numbers=none]
#program step(t).

#heuristic holds(X,V,t-1) :     holds(X,V,t). [2147483647-t, true]
#heuristic holds(X,V,t-1) : not holds(X,V,t). [2147483647-t,false]
\end{lstlisting}
% 2147483647
%
The first directive expresses that,
whenever an instance of the atom \lstinline{holds(X,V,t)} is made true by \clingo,
\lstinline{holds(X,V,t-1)} should be set to true at the
level \lstinline{2147483647-t}, % with \lstinline{true} sign,
where \lstinline{2147483647} happens to be the maximum integer supported by \clingo.
Similarly, the second directive is activated when 
an instance of \lstinline{holds(X,V,t)} becomes false,
in which case \lstinline{holds(X,V,t-1)} should be made false as well.
% and assigns the \lstinline{false} sign to \lstinline{holds(X,V,t-1)}.
%
%
Both directives associate atoms over \lstinline{holds/3} representing earlier states,
i.e., the integer taken for~\lstinline{t} is smaller,
with higher levels,
which intends to bias the search of \clingo\ towards establishing goal conditions as early as possible.
% Note that a higher level is given to atoms at earlier time instants.
%
%Initially, the atoms representing the goal are made persist backwards,
%one by one, from the last time step to the first one.
%
% Initially, the solver decides on one of the goal's \lstinline{holds/2} atoms at the last but one time step,
% then it decides to make it persist to the previous situation, and so on.
% Later, it makes persist backwards another \lstinline{holds/3} atom from the goal.
% %
% The heuristic is based on the fact that most of the times the values of the fluents persist by inertia.
%
As the experiments in Section~\ref{sec:experiments} show, 
the use of a planning heuristic often helps to improve plan search.

%The heuristic is based is two ideas.
%
%The first one is that usually the values of the fluents persist by inertia.
%
%The second one is that it is good to focus on the goals.
%}

% We have added two additional features for improving the performance of the planner.
% %
% Note that 
% when the query is posed at a time point $m$ before the end of the transition $n$, 
% the solver may spend some unnecessary efforts dealing with the atoms after $m$.
% %
% To help overcome this, the planner sets to true the atoms \lstinline{block(}$t$\lstinline{)} 
% for every $t$, $m < t \leq n$.
% %
% They can be used to constrain the part of the program after $m$. 
% %
% For example, in \plasp, the rule:
% \begin{lstlisting}[numbers=none]
% :- occurs(A,t), block(t).
% \end{lstlisting}
% enforces that after time point $m$ no action occurs. 
% %
% Hence, everything that is true at $m$ persists until $n$, 
% and the solver may easily propagate these values.
% %
% Furthermore, when this constraint is added, a second improvement is possible.
% %
% Given that the atoms at $m$ will have the same value as those at $n$, 
% we can always solve the program $P(n,n)$ instead of $P(n,m)$.
% %
% This allows eliminating the atoms \lstinline{query(}$m$\lstinline{)} once
% the transition is extended after $m$,
% simplifying the resulting program.
% %
% In our experiments we have used both of these features, which are available via command line options.

%%% Local Variables:
%%% mode: latex
%%% TeX-master: "paper"
%%% End:

% \input{gc}
\section{Translating PDDL to ASP}
\label{sec:pddl:asp}

% intro
Like its predecessors, the third series of \plasp%
\footnote{\label{note:plasp}\url{https://github.com/potassco/plasp}}
furnishes a translator from PDDL specifications to ASP facts.
These facts are then combined with ASP encodings, such as those provided in Section~\ref{sec:encodings},
and solved by an off-the-shelf ASP system, for example, by using the planner presented in Section~\ref{sec:planner}.
However, the translator integrated in \plasp~3 also goes beyond the STRIPS fragment
by supporting a range of advanced features from PDDL 3.1 \cite{ipc14www}.
Such  advanced features include conditional effects and
logical connectives as well as quantifiers in preconditions, postconditions, and goals.%
\footnote{PDDL 3.1 further allows for numeric fluents, action costs,
durative actions, preferences, and trajectory constraints,
which are not yet supported by the current version of \plasp.}
% These advanced features include conditional effects, nested expressions (in preconditions, effects, goals, and conditional effects), disjunctive preconditions, implications, universal and existential quantifiers, as well as universal effects.

To begin with, \plasp\ parses a PDDL specification into an abstract syntax tree,
which is then subject to a \emph{normalization} step in order to reduce the range
of expressions handled in the actual translation to ASP facts.
For instance, implications $\phi \rightarrow \psi$ are mapped to disjunctions $\neg \phi \vee \psi$,
and universal quantification $\forall x_1\dots x_n:\phi$ is turned into $\neg \exists x_1\dots x_n:\neg\phi$.
The latter allows for eliminating universal quantifiers, as also done by \fastdownward\ \cite{helmert06a}.
As a result, input expressions are brought into a simplified format akin to
negation normal form, except that existential quantifiers may deliberately be subject to negation.

Similar to the introduction of Tseitin variables in transforming a formula into conjunctive normal form \cite{tseitin68a},
\plasp\ further associates disjunctions and existential quantifiers occurring in its simplified format with 
\emph{derived predicates}, available from PDDL 2.2 on \cite{edehof04a}.
Derived predicates are  similar to defined fluents used in action languages $\mathcal{AL}_d$ \cite{gelinc13a} or $\mathcal{C}+$ \cite{gilelimctu03a} and, 
unlike fluents, they are not subject to inertia,
but rules for deriving them are evaluated under well-founded semantics \cite{gerosc91a} in each state.
The prerequisites of respective rule instances reflect the elements of a disjunction
or substitutions for existentially quantified variables, respectively.
As any dependency between derived predicates introduced by \plasp\ matches
an occurrence of one expression in another,
such dependencies are inherently noncircular and yield a total well-founded model in each state.
The achievement of representing disjunctions and existential quantifiers by derived predicates is
that preconditions and goals (and likewise postconditions) can be uniformly regarded as partial states
over fluents as well as derived predicates, while dedicated treatment of more complex expressions were needed otherwise.

In the final step of its translation, \plasp\ outputs
a normalized PDDL specification in terms of ASP facts.
This includes facts specifying fluents as well as derived predicates
along with their possible values, namely, \lstinline{true} and \lstinline{false}.
% The majority of the ASP facts operate on these variable--value assignments.
Moreover, actions are described by facts providing their preconditions
and postconditions, where a postcondition may in turn be subject to a condition
in order to encompass conditional effects.
Similar facts are used to express the preconditions of rules for concluding the
truth of derived predicates, where an argument specifies whether the elements of
a precondition contribute to a conjunction or disjunction, respectively.
Notably, the falsity of a derived predicate is not addressed explicitly by rules,
but rather follows ``automatically'' whenever all rules for the predicate are inapplicable.
Facts giving the values of fluents in state $\init$ as well as the goal conditions in $\goal$
then complete the factual representation of a planning task.
The detailed reference documentation of the fact format obtained by translation with \plasp\
can be found online.%
\footnote{\label{note:facts}\url{https://github.com/potassco/plasp/blob/master/doc/output-format.md}}

As an alternative to the direct translation of PDDL specifications,
\plasp\ supports the intermediate SAS format \cite{helmert06a}, obtained by
preprocessing a PDDL input with the planning system \fastdownward\
(thus following the lower branch in the workflow displayed in Figure~\ref{fig:plasp}).
On the one hand, the SAS format constitutes a modest extension to propositional STRIPS,
so that its translation to ASP facts is rather straightforward and does not involve
any sophisticated normalization step.
In fact, grounding and simplification of PDDL specifications are delegated to
\fastdownward\ in this workflow, which takes care of reducing complex expressions
to the core constructs comprised in SAS.
On the other hand, SAS brings about some particularities that are worth mentioning
and are thus addressed below.

Most notably, the SAS format features (proper) \emph{multivalued fluents},
and \fastdownward\ includes means to infer such fluents from PDDL inputs.
For instance, given a blocks world instance with $n$ blocks and Boolean fluents
such as $\mathit{on}(1,0),\mathit{on}(1,2),\dots,\mathit{on}(1,n)$ in a PDDL specification
(where $1,\dots,n$ stand for blocks and $0$ for the table),
\fastdownward\ may introduce a single multivalued fluent $\mathit{on}(1)$
with the domain $\dom{\mathit{on}(1)}=\{0,2,\dots,n\}$ for block~$1$,
as well as a corresponding fluent for each other block.
The introduction of multivalued fluents may thus reduce the overall number of fluents
and lead to a much more compact propositional representation than obtained when grounding
a PDDL specification in a naive fashion.
Beyond that, a multivalued fluent makes a functional dependency more explicit
than a group of Boolean fluents among which exactly one happens to be true in each state.
To see this, note that the choice rule in Line~9 of Listing~\ref{lst:baseline} readily expresses
that each state maps a multivalued fluent to some value in its domain,
while matching (successor) states to applied actions were required to
figure out that several Booleans cannot hold together or be all false in a state, respectively.

In addition to multivalued fluents,
the preprocessing by \fastdownward\ may infer \emph{mutex groups},
providing fluent values such that at most one of them can hold in a state.
Reconsidering a blocks world instance with $n$ blocks,
the values $\mathit{on}(1)=n,\dots,\mathit{on}(n-1)=n$ exclude each other,
and a respective mutex group makes explicit that at most one block
can be located on the block denoted by~$n$, no matter the actions leading to
a particular state.
The mutex groups inferred by \fastdownward\ are reported in the SAS format
and provide redundant/implied information that can nevertheless help to
improve plan search, as corresponding integrity constraints on
(successor) states are easy to express in ASP and readily included in the
online versions%
\footref{note:plasp}
of the encodings given in Section~\ref{sec:encodings}.

Finally, the SAS format features \emph{axiom rules} as a counterpart
for rules addressing derived predicates in PDDL.
Unlike the latter, however,
axiom rules are grouped into layers specifying an evaluation order,
rather than relying on well-founded semantics and stratification \cite{apblwa87a}
for guaranteeing a unique outcome of the rules.
In view of this difference, the current fact format\footref{note:facts} of \plasp\
distinguishes between derived predicates according to PDDL and axiom rules specified in SAS,
while PDDL and SAS inputs lead to a homogeneous factual representation otherwise.
Given that the encodings in Section~\ref{sec:encodings} focus on (multivalued) STRIPS,
so that derived predicates and axiom rules are beyond scope, % anyway,
we rely on the common fact format obtained with \plasp\ to
compare ASP-based planning with or without preprocessing by \fastdownward\
in Section~\ref{sec:experiments}.
However, we envisage to overcome the representation gap between derived predicates and
axiom rules by furnishing a common fact format for both in future versions of \plasp,
and generalizing (parallel) ASP encodings beyond the STRIPS fragment is a subject to future work as well.
Notably, a prototypical approach to encode axiom rules in ASP has been developed in
\cite{miufuk17a} and suggests functionalities for automatic axiom extraction.

Apart from translating PDDL or SAS inputs to ASP facts
(by using the \lstinline{translate} command of \plasp),
\plasp\ offers additional functionalities activated by respective commands.
These include \lstinline{normalize} in order to inspect a   normalized
PDDL specification produced by \plasp\ in PDDL syntax instead of ASP facts.
Moreover, \lstinline{check-syntax} and \lstinline{beautify} allow for
verifying whether a given PDDL specification is supported by \plasp\ or
pretty-printing it with a uniform indentation to improve readability.
Further commands will be added to future versions of \plasp\ for automated support of
PDDL requirements analysis and plan verification, % as well as the
% guess-and-check approach presented in Section~\ref{sec:planner},
among others.

Last but not least, we note that \plasp\ is implemented in C++,
pursuing a modular design geared for both efficiency and extensibility.
In particular, \plasp\ builds on top of a dedicated \lstinline{pddl} library,
which provides the parsing and normalization functionalities used in its translation.
Given that such functionalities are independent of the ASP target formalism,
the \lstinline{pddl} library might be useful for third-party planner developers as well.

\section{Experiments}\label{sec:experiments}

% \newcommand\configuration[4]{($p_{#1}$, $e_{#2}$, $h_{\operatorname{#3}}$, $c_{\operatorname{#4}}$)}

% Like its predecessor versions, the third series of \plasp%
% \footnote{\label{note:repository}Available at: \url{https://github.com/potassco/plasp}}
% provides a translator from PDDL specifications to ASP facts.
% Going beyond the STRIPS-like fragment, % of PDDL,
% it incorporates a normalization step to support advanced PDDL features such as
% nested expressions in preconditions, conditional effects, axiom rules, as well as existential and universal quantifiers.
% Moreover, \plasp\ allows for optional preprocessing by \fastdownward,
% leading to an intermediate representation in the SAS planning % exchange
% format.
% This format encompasses multivalued fluents, mutex groups, conditional effects, and axiom rules,
% which permit a compact (propositional) specification of planning tasks and are (partially) inferred by
% \fastdownward\ from PDDL inputs.
% Supplied with PDDL or SAS inputs,
% \plasp\ produces a homogeneous factual representation, %  in ASP,
% so that ASP encodings remain independent of the specific input format.%
% %
% \footnote{%
% The encodings given in Section~\ref{sec:encodings} focus on STRIPS-like planning tasks
% with multivalued fluents as well as mutex groups, where the latter have been omitted for brevity.
% In contrast to the ease of incorporating mutex groups,
% extending parallel encodings to 
% conditional effects or axiom rules is not straightforward \cite{riheni06a},
% % and left as future work,
% while sequential encodings for them are shipped with \plasp.}

To empirically contrast the different encodings and planning algorithms
presented in Sections~\ref{sec:encodings} and~\ref{sec:planner},
we ran \plasp{} on PDDL specifications%
\footnote{\label{note:bench}\url{https://github.com/potassco/pddl-instances}}
from the International Planning Competition.
For comparison,
we also include two % three 
variants of the state-of-the-art SAT planning system \madagascar\ \cite{rintanen14a},
where \mad\ stands for the standard version and % ,
\madp\ for the use of a specific planning heuristic. % , and
% \madpC\ for the additional adjustment of planning algorithm~\algb{}{}
% to algorithm~\algc{}{}, which increases the plan length exponentially rather than linearly.
The experiments were performed sequentially
on a Linux machine equipped with Intel Core i7-2600 processor at 3.8~GHz and 16~GB RAM,
limiting time and memory per run to 900 seconds and 8~GB, % respectively,\comment{JR: seconds and 8GB?}
while charging 900 seconds per aborted run in the tables below. % Tables~\ref{tab:easy} and~\ref{tab:hard}.

Regarding \plasp,
we indicate the encoding of a particular kind of plan by a superscript to the planning algorithm
(denoted by its letter), where
\encs\ stands for sequential,
\enca\ for \astep,
\ence\ for \estep,
\encc\ for \estep\ by means of acyclicity checking, and
\encr\ for relaxed \estep\ plans;
e.g., \algb{\enca}{} refers to algorithm \algb{}{}
applied to the encoding of \astep\ plans given by Listings~\ref{lst:baseline} and~\ref{lst:forall}.
The parameters of \alga{}{} and \algb{}{} are set to $n=16$ or $\gamma=0.9$, % $\gamma=\frac{9}{10}$,
respectively,
 as suggested in \cite{riheni06a}.
With all three planning algorithms, i.e., \algs{}{}, \alga{}{}, and~\algb{}{},
we fix the increment amount for increasing the plan length to five states rather
than allowing for a single additional state only,
as this accelerated increase led to generally better performance.
For example, algorithm \alga{}{} with $n=16$ initially runs the lengths
$\text{0}, \text{5}, \text{10}, \dots, \text{75}$ simultaneously.
% \comment{JR: Somewhere we have to say the values of parameters $n$ and $\gamma$ for \alga{}{} and \algb{}{}
% For $n$ I don't remember (Patrick?), for $\gamma$ it was $0.9$}
Each of the resulting algorithm/encoding combinations can optionally be augmented with
the planning heuristic described in Section~\ref{sec:planner},  % \cite{gekaotroscwa13a} %, which has been inspired by~\madp\
which is denoted by an additional subscript \heu, like in \algb{\enca}{\heu}.
Moreover, 
the superscript \encg\ stands for the guess-and-check strategy that switches from
the program~$G$ in Section~\ref{sec:planner} to the encoding of \astep\ plans
in case a sequence of sets of actions obtained with~$G$ turns out to be not \estep\ serializable,
and we below investigate the setting \algb{\encg}{\heu}
that emerged as the overall most successful combination of techniques for our benchmarks.

\begin{table}[t]
\centering
\includegraphics{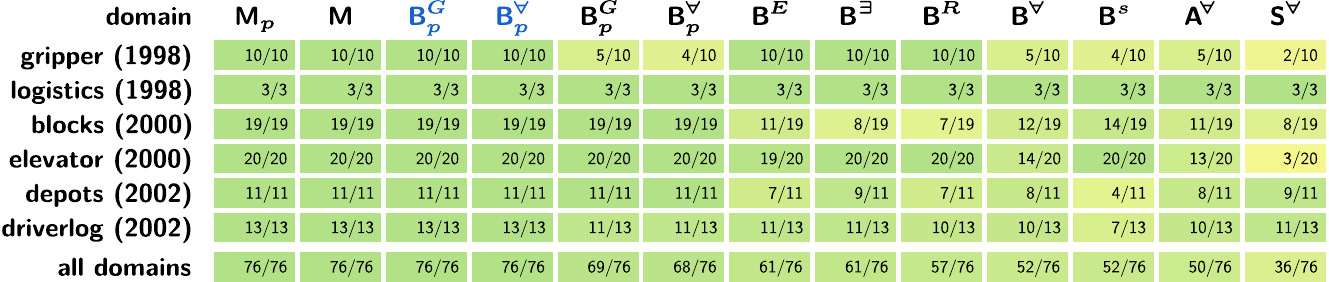}
\caption{Solved instances without preprocessing by \fastdownward\label{tab:easy}}
\end{table}
\begin{table}[t]
\centering
\includegraphics{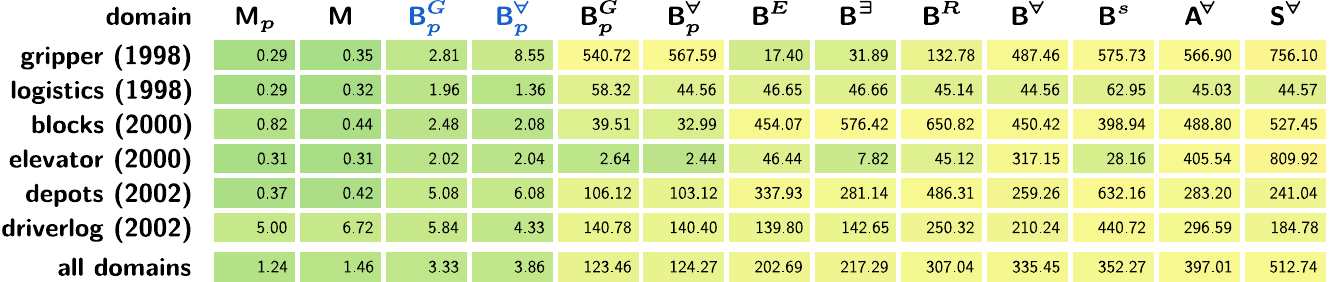}
\caption{Average runtimes without preprocessing by \fastdownward\label{tab:easyt}}
\end{table}
Tables~\ref{tab:easy} and~\ref{tab:easyt} show the numbers of solved instances and
average runtimes, for individual domains of PDDL specifications and in total,
for the two aforementioned % three 
\madagascar\ variants and % different 
\plasp\ settings that do not use preprocessing by \fastdownward.
However, for comparison we also include the two \plasp\ configurations indicated in blue,
which are based on preprocessing by \fastdownward\ and obtain their ASP facts from SAS format.
First, comparing different planning algorithms,
we observe that the sequential approach of \algs{\enca}{} falls significantly behind 
the other strategies that consider several plan lengths simultaneously.
Unlike that, the gap between \alga{\enca}{} and \algb{\enca}{},
where the latter serves as our baseline for varying the planning algorithm, encoding, or heuristic,
amounts to two more solved instances only, showing that both algorithms likewise help
to overcome costly unsatisfiable parts of the plan search.
Regarding different encodings, aiming at sequential plans with \algb{\encs}{}
works well in the inherently sequential \emph{blocks (2000)} domain as well as 
% where parallel representations cannot reduce plan length,
in the \emph{elevator (2000)} domain, although parallel representations manage to reduce plan length here.
In the \emph{depots (2002)} and \emph{driverlog (2002)} domains, however,
the performance of \algb{\encs}{} does not match parallel representations,
so that it ends up last among the encodings run with planning algorithm~\algb{}{}.
The next better \plasp\ setting, \algb{\enca}{}, improves in the latter domains
by referring to \astep\ plans, while it also exhibits particular difficulties in the
\emph{elevator (2000)} domain and does thus not solve more instances in total than \algb{\encs}{}.
The encodings of (relaxed) \estep\ plans,
utilized by \algb{\ence}{}, \algb{\encc}{}, and \algb{\encr}{},
have noticeable advantages and work especially well in the \emph{gripper (1998)} domain,
where they prove to be more effective than the encodings of sequential and \astep\ plans.
While the different implementation techniques of \algb{\ence}{} and \algb{\encc}{}
yield some performance variance, yet without a clear trend in favor of either encoding,
the extra efforts for enabling relaxed \estep\ plans with \algb{\encr}{} do not pay off
and are particularly counterproductive in the \emph{blocks (2000)} domain,
given that parallel encodings do not lead to reduced plan length  here.
The planning heuristic applied by \algb{\enca}{\heu} as well as \algb{\encg}{\heu}
constitutes an orthogonal approach
to boost plan search, which turns out to be advantageous in all but the \emph{gripper (1998)} domain
in which the encodings of (relaxed) \estep\ plans remain more successful.
However, as the two \plasp\ settings in blue that are included for comparison show,
the preprocessing by \fastdownward\ is the by far most effective way to improve plan search,
and these two settings also come close to \madagascar,
whose lead in terms of time is explained by its streamlined yet planning-specific
implementation of grounding.

% In case of Table~\ref{tab:easy}, all systems take PDDL inputs directly,
% while preprocessing by \fastdownward\ is used for \plasp\ in Table~\ref{tab:hard}.
% Note that the  tables refer to different subsets of instances,
% as we omit instances solved by some \plasp\ setting in less than 5~seconds or
% unsolved by all of them within the given resource limits.
% To give an account of the impact of preprocessing,
% we list the best-performing \plasp\ setting with or without preprocessing
% in the last row  of the respective other table.
% % exchanging the subset of instances but keeping the input format.
% % \comment{JR: For me, the last sentence ``exchanging \ldots'' makes it more complicated that without it.}
% As the \algb{\ence}{} setting of \plasp, relying on algorithm~\algb{}{} along with the (pure) ASP
% encoding of \estep\ plans, turns out to perform generally robust as well as comparable to the
% alternative implementation provided by \algb{\encc}{},
% we choose it as the baseline for varying the planning algorithm, encoding, or heuristic. % , respectively.%
% % \footnote{More benchmark results are provided at: \url{https://goo.gl/20VXD6}}

\begin{table}[t]
\centering
\includegraphics{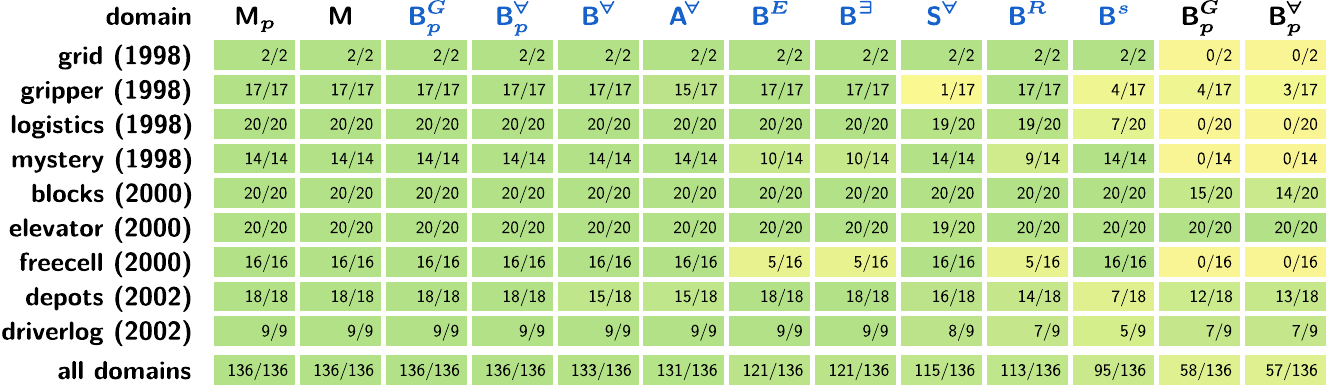}
\caption{Solved instances with preprocessing by \fastdownward\label{tab:hard}}
\end{table}
\begin{table}[t]
\centering
\includegraphics{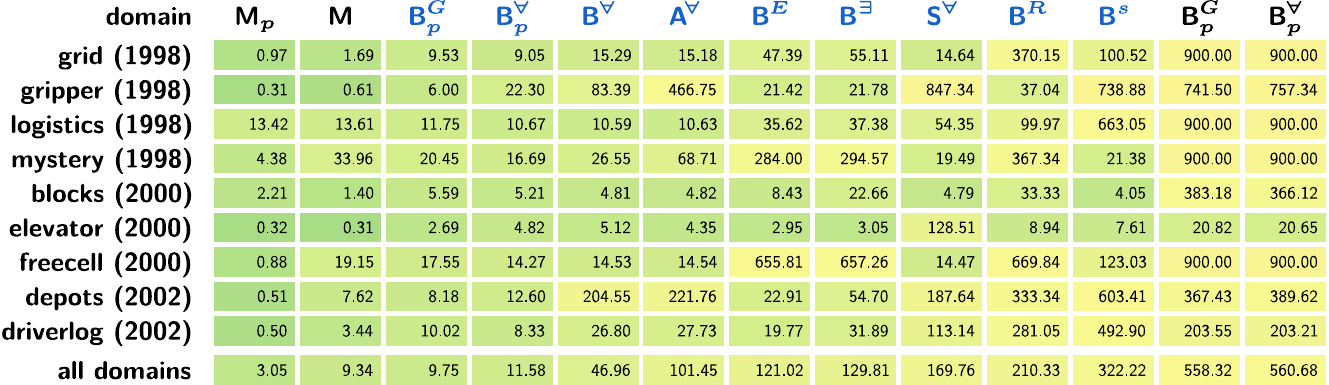}
\caption{Average runtimes with preprocessing by \fastdownward\label{tab:hardt}}
\end{table}
In the same manner as above,
Tables~\ref{tab:hard} and~\ref{tab:hardt} report numbers of solved instances and
average runtimes for \plasp\ settings that make use of preprocessing by \fastdownward,
where the \algb{\enca}{\heu} and \algb{\encg}{\heu} configurations given in black
take PDDL inputs directly and serve for comparison.
Note that the tables refer to different instances than before,
as the instances to include with or without preprocessing by \fastdownward\
were selected independently such that none of the \plasp\ settings considered in \cite{digelurosc17a}
solves an instance in less than 5~seconds, while some of these configurations
finds a plan within the given resource limits.%
\footnote{%
The requirement that some of the \plasp\ settings available in \cite{digelurosc17a}
had to finish an instance along with high computational efforts in view of frequent timeouts
are responsible for the limited numbers of domains considered in Tables \ref{tab:easy}--\ref{tab:hardt}.
Unlike that, Table~\ref{tab:rintanen} below focuses on the novel guess-and-check strategy
pursued by the \algb{\encg}{\heu} configuration and covers substantially more domains.}
The first apparent observation is that the two comparison configurations
indicated in black are outperformed by \plasp\ settings based on 
preprocessing by \fastdownward\ and translation from SAS format,
given that the introduced multivalued fluents make ground instantiations
much more compact than with Boolean fluents in ASP facts obtained directly
from the original PDDL specifications.
The \algb{\encs}{} configuration that aims at sequential plans is last among the
\plasp\ settings using preprocessing by \fastdownward, as it suffers from
greater plan lengths than needed with parallel encodings in the \emph{gripper (1998)},
\emph{logistics (1998)}, \emph{depots (2002)}, and \emph{driverlog (2002)} domains.
The next setting, \algb{\encr}{}, solves 18 instances more by referring to
relaxed \estep\ plans, while it also has particular difficulties in the
\emph{mystery (1998)} and \emph{freecell (2000)} domains, where the large number of actions
goes along with an expensive ground representation of serialization conditions.
Although the same bottleneck applies to the \algb{\ence}{} and \algb{\encc}{}
configurations, utilizing different encodings of \estep\ plans,
they manage to remain ahead of the sequential algorithm of \algs{\enca}{},
which is outperformed by the other two planning algorithms,
applied by \alga{\enca}{} and \algb{\enca}{}, in the \emph{gripper (1998)} domain.
In fact, the gap between \alga{\enca}{} and \algb{\enca}{} is again small,
and the additional incorporation of a planning heuristic in \algb{\enca}{\heu}
and \algb{\encg}{\heu} lets these two settings solve all instances
under consideration.
Similarly, the heuristic of \madp\ brings about a time advantage in comparison
to the plain version \mad\ of \madagascar, and the time difference between \madp\
and the \algb{\enca}{\heu} and \algb{\encg}{\heu} configurations of \plasp\
evolves primarily from grounding.%

\begin{table}[t]
\centering
\includegraphics{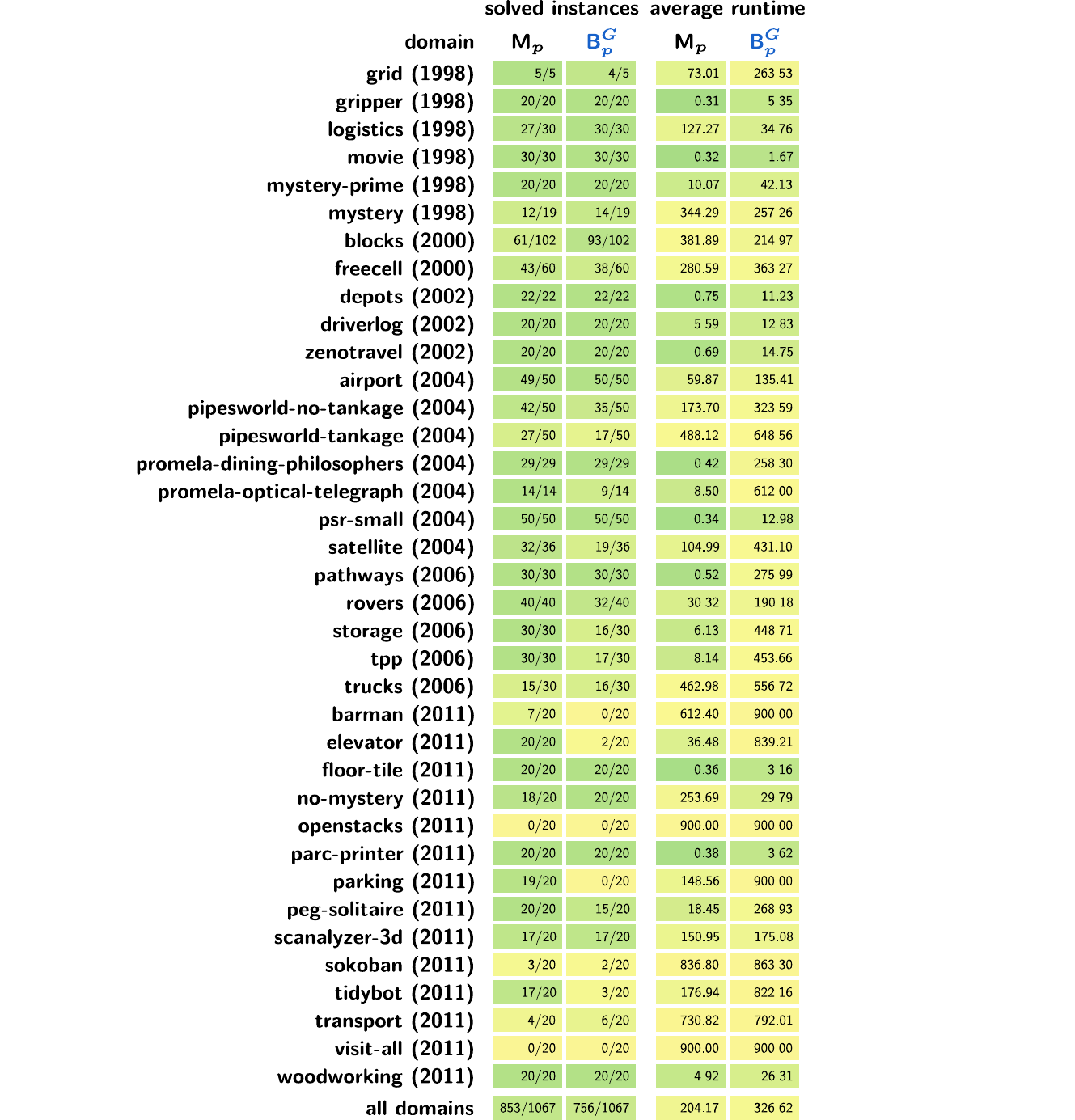}
\caption{Solved instances and average runtimes on the benchmark set by \protect\cite{rintanen12a}\label{tab:rintanen}}
\end{table}
For a broad comparison of the \madagascar\ version with planning heuristic, \madp,
and the best-performing \plasp\ setting, \algb{\encg}{\heu}, integrating
preprocessing by \fastdownward, planning algorithm~\algb{}{} along with our
guess-and-check strategy to delay serialization conditions, and a planning heuristic,
we ran both planners on the full collection of STRIPS planning tasks used in \cite{rintanen12a}.
The corresponding numbers of solved instances and average runtimes are shown in Table~\ref{tab:rintanen}.
In 15 out of the 37 domains, \madp\ and \algb{\encg}{\heu} solve the same number of instances,
\madp\ is ahead in 15 domains, and \algb{\encg}{\heu} has an advantage in 7 domains.
While the performance of both planning systems generally tends to be close,
the observed differences stem from specific implementations of grounding and heuristics,
as well as the use of an encoding of \estep\ plans by \madp,
where \algb{\encg}{\heu} switches to \astep\ plans instead
if a plan that is not \estep\ serializable is found.
In fact, in the \emph{elevator (2011)} domain, we checked that the gap of 18 solved instances
between \madp\ and \algb{\encg}{\heu} is due to the (implementation of) planning heuristic,
as \madp\ is also able to solve 16 instances more than the \madagascar\ variant \mad\
without such a heuristic.
Vice versa, \algb{\encg}{\heu} is ahead of \madp\ by 32 solved instances in the
\emph{blocks (2000)} domain, where the planning heuristic is likely to be responsible again,
given that this domain is inherently sequential so that plan length remains unaffected by
encoding differences.
Unlike that, the advantages of \madp\ in the \emph{parking (2011)} and \emph{tidybot (2011)}
domains, amounting to 19 or 14 instances solved more than with \algb{\encg}{\heu},
cannot be explained by the planning heuristic alone, but are rather related to encodings, i.e.,
the \estep\ plans obtained with \madp\ lead to reduced plan lengths in comparison to the
\astep\ plans of \algb{\encg}{\heu}.
Let us stress again that efficiency considerations regarding the ground representation of
interfering pairs of actions lead us to the guess-and-check strategy switching to \astep\
rather than \estep\ plans.
The same bottleneck has also been noted in \cite{rintanen12a}, and \madagascar\ features
a streamlined planning-specific implementation of grounding to tame the complexity of
instantiating an encoding of \estep\ plans.
While \plasp\ cannot compete with \madagascar\ regarding low-level efficiency,
its high-level approach brings the advantage that first-order ASP encodings and
general grounding facilities make it easier to prototype and experiment
with different planning algorithms and encodings.
Such flexibility has, e.g., been exploited in \cite{miufuk17a}
to implement axiom-enhanced planning, and in \cite{thielscher09a} to solve
single-player games specified in the Game Description Language (GDL; \cite{lohihascge08a})
by means of ASP planning.

% \begin{table}[t]
% \centering
% \includegraphics{figures/grounding-vs-solving.pdf}
% \caption{Grounding vs. solving times with \algb{\ence}{\heu} with and without preprocessing\label{tab:grounding-solving}}
% \end{table}

% \begin{table}[t]
% \centering
% \includegraphics{figures/solution-steps.pdf}
% \caption{Average numbers of steps of the solutions found with \algb{\encs}{} and \algb{\ence}{}\label{tab:grounding-solving}}
% \end{table}

In addition to the above experiments on PDDL domains,
we evaluated the planning algorithms \algs{}{}, \alga{}{}, and \algb{}{}
on ASP planning benchmarks, including the
\emph{hanoi-tower}, \emph{labyrinth}, \emph{no-mystery},
\emph{ricochet-robots}, \emph{sokoban}, and \emph{visit-all}
domains from recent ASP competition editions \cite{contest13a,cagemari15a,gemari17b}.
%
% For this, we took the planning problems we found in the recent ASP Competitions:
% the HanoiTower domain from 2014,
% and Labyrinth, Nomistery, Ricochet Robots, Sokoban, and Visit-all
% from 2015.
%
While the original competition benchmarks utilize non-incremental ASP encodings
along with a given maximum plan length,
we furnished incremental versions of these encodings and let the planner,
as presented in Section~\ref{sec:planner}, look for a sufficient plan length.
%
% The original encodings consist of a plain logic program, 
% which always contains a maximum plan length for each instance.
% %
% We adapted them to the program structure accepted by the planner, 
% leaving the plan length information with no effect.
%
We varied the parameters $n$ of \alga{}{} and $\gamma$ of \algb{}{} 
by using $1$, $2$, $4$, $8$, and $16$ for $n$ as well as
$0.5$, $0.75$, $0.875$, $0.9$, and $0.9375$ for $\gamma$,
which are the same values as taken in \cite{riheni06a}.
The resulting algorithm variants are denoted by subscripts,
viz.\ \alga{}{_n} for $n\in\{1,2,4,8,16\}$
and \algb{}{_\gamma} for $\gamma\in\{0.5, 0.75, 0.875, 0.9, 0.9375\}$,
where \alga{}{_1} represents the sequential algorithm~\algs{}{}.
%
% we tried algorithm \alga{}{} with $n=1 (\algs{}), 2, 4, 8,$ and $16$, 
% and algorithm \algb{}{} with $\gamma=0.5, 0.75, 0.875, 0.9,$ and $0.9375$
% (the same parameters as in \cite{riheni06a}).
%
The experiments on ASP planning benchmarks were performed sequentially
on a Linux machine equipped with Intel Core i7-6700 processor at 3.4~GHz and 32~GB RAM,
as before limiting time and memory per run to 900 seconds and 8~GB,
while taking aborted runs as 900 seconds within average runtimes.
%
% Again, we limited time and memory per run to 900 seconds and 8~GB,
%, % respectively,\comment{JR: seconds and 8GB?}
% and charged 900 seconds per aborted run. % Tables~\ref{tab:easy} and~\ref{tab:hard}.
%
%For each of them, we used an step increment \inc\ of either $1$ or $5$.

\begin{table}[t]
\centering
\includegraphics{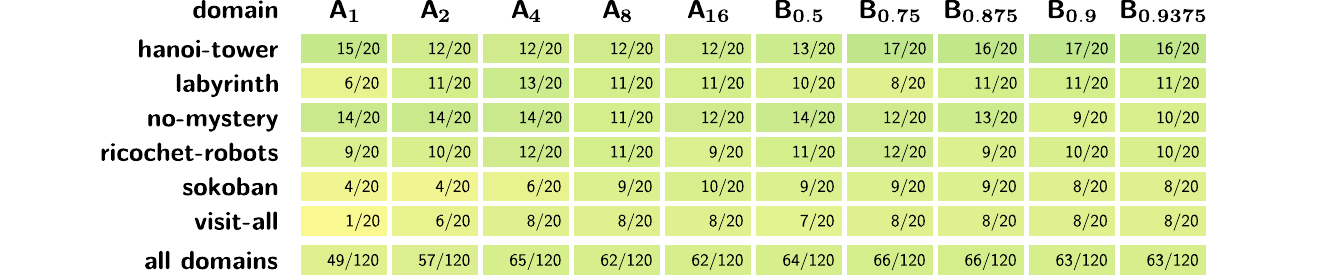}
\caption{Solved instances on ASP planning benchmarks\label{tab:asp-planning-solved-instances}}
\end{table}
\begin{table}[t]
\centering
\includegraphics{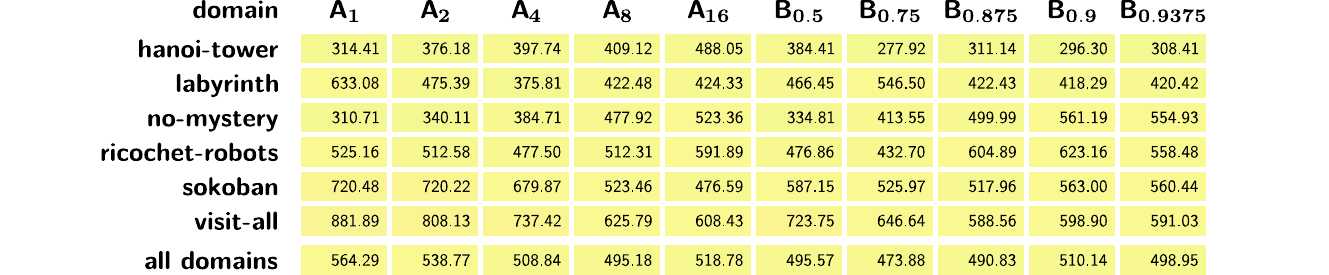}
\caption{Average runtimes on ASP planning benchmarks\label{tab:asp-planning-average-runtime}}
\end{table}
The results are summarized in Tables \ref{tab:asp-planning-solved-instances} and~\ref{tab:asp-planning-average-runtime},
showing the numbers of solved instances and average runtimes.
The overall best-performing configuration is \algb{}{_{0.75}},
but others like \algb{}{_{0.875}}, \alga{}{_4}, and \algb{}{_{0.5}} come very close,
and no setting strictly dominates over all domains.
However, it is apparent that the sequential approach of \alga{}{_1} or \algs{}{},
respectively, leads to significantly fewer solved instances than the other algorithms that
consider several plan lengths simultaneously, which particularly applies to the
\emph{labyrinth}, \emph{sokoban}, and \emph{visit-all} domains.
%
% The sequential algorithm \algs{}{} (\alga{}{}$_1$) has the worst performance,
% specially in the Sokoban and Visit-all domains,
% although in the others it performs reasonably well.
%
Comparing the length of plans found between the best-performing configuration, \algb{}{_{0.75}},
and \alga{}{_1} yields that the former include about 35 states on average,
while \alga{}{_1} leads to slightly more than 20 states only.
That is, moving on to greater plan lengths before finishing costly unsatisfiable
parts of the plan search is helpful on ASP planning benchmarks as well,
especially in order to solve instances requiring a substantial number of states
to become satisfiable.
Although we do  not show the detailed results here,
let us note   that successively incrementing the plan length by five states at once,
rather than adding a single state only, likewise leads to better performance on inputs 
obtained by translation from PDDL and direct ASP encodings of planning problems.
Given that similar solving strategies turn out to be advantageous in both cases,
developing metaencoding approaches that are applicable to incremental ASP encodings, in order
to condense stable models in the same fashion as parallel representations do for
sequential plans, may be a promising way to further speed up multishot ASP solving in the future.%
%
% Comparing in more detail these two configurations, 
% we observe that the average length of the plans found by \algs{}\ was $21.6$ steps,
% while for \algb{}{} it was $35.6$.
% %
% In particular, for the instances solved by \algs{}{},
% the average plan length of \algb{}{} 0.75 was $28.4$ steps,
% while for the others it was $53.4$.
% %
% This shows that \algb{}{} finds plans of longer length, 
% and that this ability is crucial to solve the instances not solved by \algs{}{}. 
%
% In addition, we note that the plan lengths of \algb{}{} 0.75 for Sokoban and Visit-all 
% are the largest of all domains, with $46.1$ and $74.4$ steps, respectively, 
% compared to the $38.5$, $9.4$, $24.2$ and $26.7$ for Hanoi, Labyrinth, Nomistery and Ricochet Robots, respectively.
% %
% This could partly explain the differences we observed in those domains.
%
% We also ran the benchmarks using a step increment of $1$ instead of $5$,
% but for all configurations the performance deteriorated.
% %
% In summary, the results indicate that the planning algorithms may speed up ASP planning.

%%% Local Variables:
%%% mode: latex
%%% TeX-master: "paper"
%%% End:

%%% Local Variables:
%%% mode: latex
%%% TeX-master: "paper"
%%% End:

\section{Conclusion}\label{sec:discussion} % \section{Summary}\label{sec:discussion}

We presented the key features of the new \plasp~3 system,
providing a translator from PDDL specifications to ASP facts
along with a multishot ASP planner based on \clingo.
While the ASP metaencodings in Section~\ref{sec:encodings}
focus on STRIPS-like planning tasks,
\plasp's translator component also supports a range of advanced features from PDDL 3.1
as well as the intermediate SAS format.
Moreover, its planner can be applied to any incremental ASP encoding
and thus to dynamic domains at large.
% \plasp's major components, such as the encodings as well as its planner, can be applied to dynamic domains at large.
As the experiments in Section~\ref{sec:experiments} show,
the resulting general-purpose approach comes close in performance
to the state-of-the-art SAT planning system \madagascar,
where differences are mainly due to the specific implementations of grounding and heuristics.
In particular,
general grounding techniques can constitute a bottleneck on
PDDL domains involving a large number of actions or fluents,
while dedicated preprocessing as provided by \fastdownward\ 
helps to make the propositional representation of planning tasks more compact.
The major benefit of the high-level approach of \plasp\ is that first-order
ASP encodings and general grounding means facilitate prototyping and experimenting
with different planning algorithms and encodings.
As future work, we intend to
generalize our ASP encodings of parallel plans beyond the STRIPS fragment of PDDL,
in order to support the advanced PDDL and SAS features mentioned in Section~\ref{sec:pddl:asp},
e.g., conditional effects and derived predicates or axiom rules, respectively.
Whether parallel representations of plans can be further adopted to express
the stable models of arbitrary incremental ASP encodings more compactly
is also an interesting open question that may be addressed in the future.

% \begin{itemize}
%   \item The original planning as satisfiability approach dates back more than two decades  \cite{kausel92b}.
%   \item Over the years planning encodings have been improved by utilizing ideas from other planning research, such as planning graphs, SAS+ representation, and various forms of macro-operators and abstractions \cite{kausel99a,chhuxizh09a,rogrphsa09a,siddim10a,huchzh10a,ridofrba16a}.
%   \item Apart from classic STRIPS-like planning problems, satisfiability 
%   encodings have been also used for more advanced planning problems, such as 
%   temporal planning or problems with preferences \cite{juhsmc12a,rangha15a} 
%   \item Recently, the trend of applying SAT technology to richer planning languages has led to SMT-based planning for modelling temporal problems, features of PDDL+ such as continuous change \cite{shidav05a,cafoloma16a}, as well as the integration of motion and task planning in robotic applications \cite{husotalizhwaga14a,dakichka16a}.
% \end{itemize}

%%% Local Variables: 
%%% mode: latex
%%% TeX-master: "paper"
%%% End: 

% \paragraph{Acknowledgments}
\medskip\par\noindent\emph{Acknowledgments.} \
This work was partially funded by DFG grant SCHA 550/9.
The second author was supported by
KWF project 28472,
cms electronics GmbH,
FunderMax GmbH,
Hirsch Armbänder GmbH,
incubed IT GmbH,
Infineon Technologies Austria AG,
Isovolta AG,
Kostwein Holding GmbH, and
Privatstiftung Kärntner Sparkasse.
We are grateful to the anonymous reviewers for their helpful comments.

%%% Local Variables: 
%%% mode: latex
%%% TeX-master: "paper"
%%% End: 

% \bibliographystyle{acmtrans}
% \bibliography{lit,akku,procs} % https://svn.cs.uni-potsdam.de/svn/reposWV/Papers/bibfiles/trunk

%%% Local Variables: 
%%% mode: latex
%%% TeX-master: "paper"
%%% End: 

\end{document}